\documentstyle[aps,prb,epsf,rotate]{revtex}

\newcommand{\ignore}[1]{}
\begin{document}
\twocolumn[
\hsize\textwidth\columnwidth\hsize\csname @twocolumnfalse\endcsname
%\draft
\title{ Continuum bound states as surface states }
\author{D. W. L. Sprung, P. Jagiello, J. D. Sigetich }
\address{
  Department of Physics and Astronomy, McMaster University\\
  Hamilton, Ontario L8S 4M1 Canada
}
\author{and \, J. Martorell}
\address {Dept.
  d'Estructura i    Constituents de la Materia, Facultat F\'{\i}sica,\\
   University of Barcelona, Barcelona 08028, Spain
}
\date{\today}
\maketitle

\begin{abstract}
We discuss the relation between continuum bound states (CBS) 
localized on a defect, and surface states of a finite periodic 
system. We use the transfer matrix method to model an experiment of 
Capasso, and find all continuum bound and anti-bound states. We 
compute the rate for intra-subband transitions from the ground state 
to the CBS and derive a sum rule. Finally we show how to improve the 
confinement of a CBS while keeping the energy fixed. 
 \end{abstract}
\narrowtext
]

\section{Introduction}

In 1992, an experiment of F. Capasso et al. demonstrated\cite{Cap92} 
the existence of well localized continuum bound states (CBS) in a 
semiconductor superlattice consisting of one thick quantum well 
surrounded on both sides by several GaInAs-AlInAs well/barrier layers 
constructed to act as $\lambda/4$ Bragg reflectors. As 
suggested by Lenz and Salzman \cite{Lenz90}, the central well was 
made double the width of the lattice wells, to act as a 
$\lambda/2$ Fabry-Perot resonator. Subsequently, Weber \cite{Weber94} 
studied the existence of such states using the transfer matrix 
method. Among other things, he showed that the Bragg condition need 
not be very well satisfied for a confined state to exist. 
B. Sung et al. \cite{Sung96} have also studied above threshold 
confined states, in a different material system GaAs/AlGaAs. 
Radovanovi{\'c} et al. \cite{Rado99} have written several papers 
describing mainly numerical methods for tailoring a heterostructure 
to produce CBS with equal energy spacing, to facilitate second 
harmonic generation. Finally, Xue-Hua Wang, et al. \cite{Wang99} have 
discussed the parity sequence of sub-threshold bound states localized 
on a defect, and the transition rates between them, but neglecting 
the variable effective mass. 

In this paper we provide further insight into the 
phenomenon of CBS by relating them to surface states, whose existence 
was elucidated by Shockley in a deservedly famous paper \cite{Surf}. 
An infinite periodic system, illustrated by line (c) of Fig. 
\ref{fig1}, allows Bloch states with the periodicity of the lattice. 
If the system is truncated on one side, or on both sides, (line (b)) 
then one can discuss scattering states with energies above threshold, 
and bound states below threshold. The transfer matrix method is well 
adapted to discuss such a periodic system. For convenience we will 
use the notation of our previous papers \cite{SWM92,SSWM00}. 

    Among the bound states of a finite periodic array are the Bloch states 
whose amplitude is spread more or less uniformly over the lattice, 
and the surface states whose density is confined to the ends. The 
former usually occur in the allowed energy bands of the infinite 
lattice, while the surface states necessarily occur in the forbidden 
bands. Their wavefunctions decay exponentially outside the array and 
like $\pm e^{-\theta}$ from cell to cell inside, where $\theta$ is the 
imaginary part of the Bloch phase. 

Another way to truncate an infinite lattice is to cut it in the 
middle and pull the two halves apart. This introduces a defect 
(line (d) of Fig. \ref{fig1}) which may be either a well or a 
barrier. As emphasized by Weber \cite{Weber94}, the condition for a 
state localized in the central gap of an infinite lattice is that the 
boundary condition at the edge of the defect matches to a decaying 
Bloch eigenstate of the unit cell: that is, the wave function will 
decay by $\pm e^{-\theta}$ from cell to cell. There are also 
anti-bound states, where the match is to the growing eigenstate. In 
either case, the Bloch phase must be complex, so such states exist 
only in the forbidden Bloch zones (FZ). 

In section 2, we set up the (generalised) transfer matrix for a 
system with position-dependent effective mass. In section 3 we apply 
it to the Capasso experiment. We determine the width of 
the central well/barrier to provide a CBS at a desired  energy. In 
sec. 4 we discuss the relation between these states and states in a 
box, illustrated in frames (a) and (f) of Fig. \ref{fig1}. In Sec. 5 
we show how to compute the transition rate from the ground state to 
continuum states in the neighbourhood of the CBS, and derive a sum 
rule which implies that the total strength depends very little on the 
number of cells involved. Finally, in Sec. 6 we propose new 
experimental configurations that improve 
the confinement of a CBS while keeping the energy fixed. 

\section{Transfer Matrix}

Consider solutions of the Schr\"odinger equation with a variable 
effective mass 
 \begin{equation}
-{\hbar^2\over 2} {d\over {dx}}\left( {1\over {m(x)}} {d\over {dx}} \psi\right) 
+ (V(x) -E) \psi = 0
\label{eq:t1} 
\end{equation}
At a discontinuity of the potential or of the effective mass, both 
$\psi(x)$ and $\psi'(x)/m(x)$ are continuous.
Let $v(x)$ and $w(x)$ be any two independent solutions at the same $E$. 
Then the modified Wronskian 
 \begin{eqnarray}
 w(x) \frac{v'}{m(x)} - v(x) \frac{w'}{m(x)} &=& C  
\label{eq:t2}
\end{eqnarray}
takes a constant value. 
We choose $C = 1$ by suitable normalization. 

For convenience we will discuss the situation where the entire system 
has reflection symmetry. Then it is sufficient to consider solutions 
only for $x > 0$ and reflect them in the origin. In many cases, 
including the specific examples discussed by Weber, Capasso and Wang 
et al., it is possible also to choose the unit cell to have 
reflection symmetry about its midpoint. We will do this when it aids 
in the analysis. 

We factorise $m(x) = m_e m^*(x)$ into the bare mass $m_e$ and the 
dimensionless $m^*$. The constant $\hbar^2/2m = 3.81$ eV 
$\AA^2$ converts from energy to length units. In Weber's model both 
the potential and the effective mass are piece-wise constant 
functions, but the method is valid even if $V(x)$ and $m^*(x)$ vary 
continuously within the cell. 

Consider a solution $v(x)$ which has value 1, slope 0 at the left 
edge of the unit cell, and another solution $w(x)$ which has value 0 
and slope $m^*(0)$ there. The transfer matrix for the unit cell of the 
lattice is then 
 \begin{eqnarray}
W(0 \to d) &=& \pmatrix{v & w \cr v'/m^* & w'/m^* }   
\quad {\rm with } \nonumber \\ 
\pmatrix{\psi \cr \psi'/m^*}_{d} &=& W(0 \to d) \pmatrix{\psi \cr 
\psi'/m^*}_{0} \, . 
\label{eq:t3}
\end{eqnarray}
Here the solutions $v, w$ without argument are evaluated at the cell 
edge, $x = d-0$, and $\psi$ is an arbitrary solution. In a periodic 
system, $W(d)$ depends only on the length of the cell, not its 
position. Since $\psi$ and $\psi'/m^*$ are continuous at interfaces, 
to move one interval further to the right one simply multiplies again 
by the appropriate transfer matrix. Any discontinuity in the 
derivative is automatically taken into account. 

Eigenvalues of the transfer matrix satisfy
 \begin{eqnarray}
\lambda^2 - 2\cos \phi \lambda + 1 &=& 0  \quad {\rm where} 
\nonumber \\
2 \cos \phi \equiv {\rm Tr} \,W &=& v +  w'/m^*  \, , 
\label{eq:t4}
\end{eqnarray}
and when the Bloch phase $\phi$ is a real angle, they are $\lambda = 
e^{\pm i\phi}$. Generally, raising the energy from the potential minimum, one 
is in a forbidden band where $|\cos \phi| > 1$. In this region of 
energy, $\phi = i\theta$ is imaginary. Following this, the first 
allowed band occurs within which $\phi$ increases from zero to $\pi$. 
Then in the next forbidden band, with $\cos \phi < -1$, $\phi = \pi + 
i\theta$ becomes complex. In the $p$'th forbidden band, $\phi =  p\pi 
+ i\theta$ and the eigenvalues are $\lambda = (-)^p e^{\pm \theta}$. 

\subsection{Surface States}
Because we have assumed that parity is a good quantum number, states 
of the whole system will have either even or odd parity. 
Suppose that the infinite array is truncated so that there are $N$ 
cells to right of the origin as in Fig. 1, frame (b). Then the 
condition for a bound state is that the wave function at the right 
edge of the array matches to a decaying solution outside. (Here we 
suppose constant potential outside, but that can be changed 
trivially.) 
 \begin{eqnarray}
\pmatrix{\psi \cr \psi'/m^*(d)}\Bigr|_{Nd} &=& W^N  \pmatrix{\psi \cr 
\psi'/m^*(0)}\Bigr|_{0} 
\label{eq:t5}
\end{eqnarray}
For an even bound state, $\psi = v(x)$, on the right hand side $W^N$ 
acts on $(1,0)$ while for an odd state $w(x)$ it acts on $(0, 1)$. 
This gives the log-derivative 
 \begin{eqnarray}
\frac {\psi'}{m^*(d)\psi}\Bigr|_{Nd} &=& 
\frac{W^{(N)}_{2s}}{W^{(N)}_{1s}}
= \frac{-\kappa}{m^*_{out}}
\label{eq:t6}
\end{eqnarray}
where $s=1 \, (2)$ for even (odd) states, and $m^*(d)$ is the effective 
mass inside the edge of the last cell, while $m^*_{out}$ is the 
value outside, and $E = V_{out} -\hbar^2 \kappa^2/(2 m m^*_{out})$. By 
construction, the $W$-matrix is real, so the energy $E$ must lie 
below the external potential $V_{out}$. On the other hand, a 
surface state can exist only when $E$ is in a forbidden zone, with 
complex $\phi= p\pi + i\theta$. In such a zone \cite{SWM92}, 
 \begin{eqnarray}
W(Nd) &=& W^N(d)  \nonumber \\
&=& (-)^{(N-1)p} \bigl[ \frac{\sinh N\theta}{\sinh \theta} 
W - (-)^p\frac{\sinh (N-1)\theta}{\sinh \theta} \bigr] 
\nonumber \\ 
\label{eq:t7}
\end{eqnarray}
Eqs. \ref{eq:t6}, \ref{eq:t7} allow one to search for energies 
where surface states occur.

\subsection{Continuum Bound States}
Suppose that the infinite periodic array is cut at the origin and 
an extra well of width $2c$ is placed between the two sections, as in 
frame (d) of Fig. \ref{fig1}. Let $T(c)$ be the transfer matrix (as in 
eq. \ref{eq:t3}) that takes the wave function from the origin to 
$c$. Its columns are the even and odd parity solutions within the 
central well. In order for a CBS to exist, the first (or second) column 
of $T(c)$ must match to a decaying eigenstate of the unit cell of the 
semi-infinite array to the right. In other words, one of the columns 
of $T(c)$ must satisfy the eigenvalue equations for $W(d)$ \cite{Weber94}: 
 \begin{eqnarray}
(W(d) - \lambda I) \pmatrix{\psi(c) \cr \psi'(c)/m^*} &=& 0 
\label{eq:t8} 
\end{eqnarray}
 \begin{eqnarray}
\frac{\psi'(c)}{m^*(c) \psi(c)} &=& \frac{\lambda - W_{11}(d)}{W_{12}(d)}
\nonumber \\
&=& \frac{W_{21}(d)}{\lambda - W_{22}(d)}
\, . 
\label{eq:t9}
\end{eqnarray}
Either of these equations can be used to search for CBS. If they are 
satisfied with real $\lambda$ being the smaller (larger) eigenvalue, then 
a CBS (or an anti-bound state, (ABS)) exists at that energy. The only 
difference between them and surface states is the numerical value of 
the boundary condition that has to be satisfied where the lattice 
meets the defect, or the surface. 

\section{ Capasso-Weber example}
The array constructed by Capasso can be modelled as a sequence of 
potential wells of width $a= 16 \AA$, depth $V_w=0$ and barriers of 
width $b = 39 \AA$, height $V_b = 500$ meV. The energy-dependent 
effective mass in each layer is given by \cite{NMK87} 
 \begin{eqnarray}
m^*_w &=& 0.043 ( 1 + (E-V_w)/E_w ) \nonumber \\
m^*_b &=& 0.073 ( 1 + (E-V_b)/E_b ) 
\label{eq:t10}
\end{eqnarray}
where $E_w = 0.88$ eV, and $E_b = 1.49$ eV are the effective band 
gaps of the InGaAs wells and AlInAs barrier materials. 

In this example, the potential is piece-wise constant, so the 
transfer matrix can be constructed from factors of the type 
 \begin{eqnarray}
T(c) = \pmatrix{ \cos kc & m^*_c \sin kc/k  \cr 
                -k \sin kc/m^*_c & \cos kc }
\label{eq:t11}
\end{eqnarray}
where $k^2 = 2 m_e m^*_c (E-V_c)/\hbar^2$ is the wave number inside the 
layer and $m^*_c$ is the effective mass there. In contrast to Weber, 
we find it convenient to define the unit cell as a half-well of width 
$a/2$ on each side of a barrier of width $b$: then 
 \begin{eqnarray}
W(d) = T(a/2) T(b) T(a/2)
\label{eq:t12}
\end{eqnarray}
is the transfer matrix of a symmetric unit cell, and has both 
diagonal elements equal to $\cos \phi$. Panel (a) of Fig. \ref{fig2} 
shows a part of the array.  

In Fig. \ref{fig3}, we show the eigenvalues in the forbidden zones. 
To guide the eye, they are connected by straight lines across the 
allowed zones. There is an even index forbidden zone up to 307 meV in which 
the eigenvalues vary enormously. The odd FZ from 387 to 641 meV shows 
a smaller but still appreciable range of variation. This is reduced 
still further in the third FZ from 881 to 960 meV. The smallest 
values of $|e^{-\theta}|$ occur near the middle of the forbidden 
zones. 

The transfer matrix from the origin to $ c + Nd$ is $ W^N(d) T(c)$. 
For any non-zero $c$ the central well constitutes a defect in the 
superlattice. 
If $N \to \infty$, the argument of Sec. 2B applies, and the 
wave function will only be localized near the origin when one of the 
columns of $T(c)$ is an eigenvector of $W(d)$ with the decaying 
eigenvalue $\pm e^{-\theta}$. The left hand side of eq. \ref{eq:t9} 
(cf. eq. \ref{eq:t11}) is either $-k_w \tan k_wc /m^*_w$ (for an even 
state), or $k_w \cot k_wc /m^*_w$ (odd state) . But the identity 
$\cot (x + \pi/2) = - \tan (x)$ means that the solutions for odd 
states can be found simply by adding $\pi/2$ to the value of $k_w c$ 
of an even state solution. Given any solution, another one which 
differs only by the number of nodes in the central well, can be 
obtained by adding $\pi/k_w$ to $c$. Hence, it is the {\it 
differences} in width of the central well that go by 
half-wavelengths, not the whole width. 

By solving 
 \begin{eqnarray}
\tan k_w c &=& \frac{ W_{11}(d) - \lambda}{W_{12}(d)} \frac{m_w}{k_w}
\nonumber \\
&=& \frac{W_{21}(d)}{ W_{22}(d) - \lambda} \frac{m_w}{k_w}
\label{eq:t13}
\end{eqnarray}
we obtain the values of $c$ in Fig. \ref{fig4} where we 
plot the solutions from the principal branch of the arctan function. 
In the first (even) FZ, lines 2 and 4 reach the $c=0$ axis at the 
beginning of the allowed zone, at 641 meV. In our convention, $c=0$ 
makes the central well have total width $a = 16 \AA$, the same as in 
the infinite SL, so one has a periodic system in which no confined 
state exists. That is why the loci of CBS or ABS terminate on the 
line $c=0$ at either end of a forbidden zone. 

The horizontal dotted line at $c=8 \AA$ picks out the odd CBS state at 
563 meV found by Weber and Capasso. In this case, the central well is 
double the width of the wells in the lattice. Very close by at 577 meV 
is an even ABS (line 3). In the next odd-parity  FZ there is an odd parity 
CBS at about 890 meV. 

{}Fig. \ref{fig5} allows one to read off the energies of the CBS 
(ABS) for any choice of width $c$. Line 2 is just line 1  augmented 
by $\pi/(2 k_w)$; similarly for lines 4 and 3.  Negative values of 
$c$ down to $-a/2$ are meaningful, as they simply shrink the width of 
the central well down to zero. The straight line segments across the 
allowed zones are drawn just to connect related solutions. 

Because the choice of a flat central potential leads to the tangent 
function in eq. \ref{eq:t13}, one can invert to obtain $c$ directly. 
For a more complex potential one would have to search for solutions 
of eq. \ref{eq:t9}, but nothing would change in principle. 
Any potential that produces the same boundary condition at $c$ gives 
the same solution, with only the details of the wave function between $\pm 
c$ changing. For example, Wang et al. \cite{Wang99} used a two-segment 
well/barrier combination as their central well.  Its boundary value 
still varies between $\pm \infty$, so one must expect similar 
solutions, albeit at slightly different energies. 

As explained by Shockley \cite{Surf}, (see also \cite{SSWM00}) a 
unit cell which is more attractive at the edges than in the centre, is just 
the type which favours the occurrence of surface states. Compared to 
a well of constant average depth, the energies of even states are 
raised and those of odd states, lowered. This leads to band mixing in 
the forbidden zones, making it possible for the decaying eigenstate 
of the cell to match to a decaying exponential with a positive log 
derivative of $\psi$ at the outer edges. In the Capasso device, it is the 
odd state in the central well which has the positive log derivative 
where it matches to the lattice on the right, making the CBS.  Viewed 
from the lattice, at the left edge the match is to a fast decreasing wave 
function in the central well, which must vanish at the origin.  

To illustrate the close relationship between CBS and surface states, 
in Fig. \ref{fig6} we show in the upper panel the right side of a six cell 
lattice with the central well of width $2a$, having the CBS at 563 
meV. In the lower part we show a three cell lattice with a surface 
state at the left edge decaying into the potential barrier, chosen so 
the slope is exactly the same for both wave functions. The only 
difference is that one state goes to zero at the origin to make an 
odd parity wave function, while the other decays exponentially; 
within the lattice they are identical.

As a second example, we replaced the central well by a central 
barrier, (see Fig. \ref{fig2}b) similar to the system proposed by 
Lenz and Salzman \cite{Lenz90}. In this case the unit cell is taken 
to be a barrier of width $b/2$ on either side of a well of width $a$. 
The diagonal elements of $W(d)$ are interchanged but its trace is 
unchanged, so the band structure is identical. In Fig. \ref{fig7} we 
show the corresponding results for the extra barrier-width $c$. As 
before, $c=0$ corresponds to a periodic system and no CBS is 
possible, as this point comes at the edge of an allowed band. The 
central barrier pushes up the energies of  even parity states, so now 
an even-parity CBS is found in the neighbourhood of 550 meV, with $c 
\sim 20 \AA$. There is a nearby odd-parity ABS if $c>17 \AA$. 

To see whether the thicknesses chosen by Capasso were optimal, we 
varied $b$ and $a$. For fixed well 
width $a$, the area enclosed by the eigenvalues of the odd FZ in Fig. 
\ref{fig3} expands continually as $b$ is increased, even to $200  
\AA$. The lowest absolute value of the smaller eigenvalue $|\lambda_-
|$ gets smaller and smaller: see Table  \ref{tx1}. However, this 
simply means that the wells become progressively decoupled as the 
distance between them gets larger. Similarly, for fixed barrier width 
$b= 39 \AA$,  $|\lambda_-|$ decreases steadily as the well width 
increases, while the energy of the CBS moves down towards threshold. 
At $a = 20 \AA$ it becomes a true bound state below threshold. For 
$a = 19 \AA$, $|\lambda_-| = 0.2508$ which is almost twice the rate 
of decay of probability (which depends on $|\lambda|^2$) 
from cell to cell within the lattice, as 
compared to $a=16 \AA$. The energy of the CBS is then at 506 meV. The 
values $k_w a = 1.810$ and $k_b b = 0.426$ are even further from the 
Bragg condition than for the original parameter set. This confirms 
Weber's finding that a high degree of confinement does not require 
exactly satisfying the Bragg condition. 

For a central barrier, the behaviour with increasing barrier width is 
similar to what was found above for a central well: see Table \ref{tx2}. 
Here it is an even parity CBS that falls in energy from 
778 to 503 meV as $b$ increases from 20 to 100 $\AA$. But again, this 
is a result of decoupling of the wells. For fixed $b = 39 \AA$, there 
is a shallow minimum of $|\lambda_-| = 0.351$ at $a = 17 \AA$ and the 
energy of the CBS is at 574 meV. Again, $k_w a = 1.757$ and $k_b b = 
1.505$ are far off the Bragg condition. 

However, the conclusion is very different if both $a$ and $b$ are varied 
while keeping the energy of the CBS fixed. An example is shown in 
Table \ref{tx3}. The minimum value of $|\lambda_-|$ 
is obtained with both $k_w a$ and $k_b b = \pi/2$. It should be 
expected because, once we fix the energy of the state, the effective 
mass is also fixed. Then the optimization of $k_w a$ and $k_w b$ 
proceeds exactly as for an energy-independent Kronig-Penney 
potential, for which the Bragg condition is optimal, as one can 
easily show analytically. 

\section{Relation to states in a box}

Kalotas and Lee \cite{KL95} considered the states obtained by 
enclosing a finite number of cells between infinite walls. (See frames 
(a) and (f) of Fig. \ref{fig1} for an illustration.) This discretizes 
the continuum, so all states become discrete. Well localized states 
that decay quickly enough will be scarcely affected by the walls. 
States spread over the whole lattice will become a discrete set 
maintaining a similar character. An ABS whose magnitude grows away 
from the origin will be squeezed against the walls of the box.

In order for the finite $N$ system to agree with Weber's, we 
add an extra half-well or barrier at the right edge of the array. 
Fig. \ref{fig8} shows an example where we have taken $N=10$ cells on 
each side of the origin. This is to be compared with the CBS at $E = 
563$ meV of the $c=8 \AA$ example of Capasso and Weber. Even with just 
three cells on either side, the state is hardly shifted from its 
position in the infinite array. 

In Fig. \ref{fig9}  we show the spectrum of box states as a 
function of $N$, again for the $c=8 \AA $ central well case. The energies of 
the single-cell states change little as more cells are added. The new 
states that appear fill up the allowed bands. To understand this, it 
is convenient to consider a system with a hard wall at the origin, 
then $N$ identical cells, followed by a hard wall at the right. 
The allowed wave functions are those that vanish at the origin 
(odd-parity states of the symmetric system), and the hard-wall boundary 
condition at the right edge is $\psi(x = Nd) = 0$. In view of eq. 
\ref{eq:t5}, this requires that the element $W_{12}(Nd) = 0$.
Since in an allowed band 
 \begin{eqnarray} 
W^{N}(d) =  \frac{\sin N\phi}{\sin \phi} W(d) - 
\frac{\sin (N-1)\phi}{\sin \phi} I \, ,
\label{eq:t14}
\end{eqnarray} 
this can be written 
 \begin{eqnarray} 
W^{(N)}_{12} = 0 = \frac{\sin N\phi}{\sin \phi} W_{12}(d) 
\label{eq:t15}
\end{eqnarray} 
This shows that 
bound states can occur in either of two ways. First, as single-cell 
bound states, where the second factor vanishes. These states have 
$N$ nodes, and the wave function vanishes at every cell boundary. 
Alternatively, the combinatorial factor $\sin N\phi/\sin \phi$ might 
vanish, and in an allowed band there are $N-1$ such states where 
$N\phi = m\pi$, $m = 1, 2, ... N-1$.  The single-cell state may occur 
in a forbidden zone, but the others can only occur for real $\phi$, 
in an allowed band. For non-zero $c$ one has to multiply $W^N(d)$ 
from the right by the additional transfer matrix $T(c)$ for the 
central well, and then the simple factorization won't be exact. In 
practice the states tend to remain in the allowed band all the same. 

Fig. \ref{fig9} provides another example of the similarity of CBS and 
surface states. In the upper panel (a), the CBS lies in the middle of 
the first of the first forbidden zone, while in the lower panel there 
are two such states, one derived from allowed band $\alpha$ and the 
other from band $\beta$.

\section{Transition Rates} 

Introducing the vector potential into the Hamiltonian eq. \ref{eq:t1} 
leads to the excitation operator 
 \begin{equation}
\frac{e A}{2c} [ p \frac{1}{m(x)} + \frac{1}{m(x)} p] \equiv \frac{-
i \hbar eA}{2m_e c} S
\label{eq:t16}
\end{equation}
with dimensions of energy. 
In defining the operator $S$, (dimensions inverse length) we have 
factored out the bare electron mass, leaving only the dimensionless 
effective mass ($m^* \sim 0.06$) inside. The vector potential $A$ is 
assumed to be a function of $x$, so it commutes with the mass. By 
invoking the Coulomb gauge we make it commute with the momentum as 
well. 

According to the Golden rule, the transition rate is 
 \begin{equation}
w_{if} = \frac{2\pi}{\hbar}\left(\frac{eA \hbar}{2 m_e c}\right)^2
|<\Psi_f| S |\Psi_i>|^2 \rho(E_f) \, ,
\label{eq:t17}
\end{equation}
where $\rho(E_f)$ is the density of final states. 
The factors before the matrix element have dimensions length squared 
energy per second, and these are omitted from our calculations. 
The matrix element squared times the density of states is therefore 
(energy length-squared)$^{-1}$, and this is what we plot in Fig. 10 
for example. After integrating over energy, we have units $\AA^{-2}$ 
for the total strength. 

In the Capasso experiment, the ground state has even parity, so its 
derivative is odd, and transitions are allowed only to odd parity 
excited states. Also the lattice is finite rather than infinite, so 
the transitions are to states in the continuum. In the neighbourhood 
of the CBS, the continuum wavefunction has a large normalization 
inside the central well, and this causes the transition rate to peak 
at or near this energy. 

We again consider the case of a central well of half-width $c$ 
surrounded by $N$ cells on each side. But for convenience, 
in this section we follow Weber by defining $W(d) = T(a)T(b)$ as having a 
barrier of width $b$ on the left followed by a well of width $a$. 
Then $N$ will be the number of wells additional to the central well. 
The odd parity excited state with Dirac delta-function normalization 
has wave function 
 \begin{eqnarray}
\Psi_f(x) &=& B_0 \sin (k_w x) \quad , \quad |x| < c \nonumber \\ 
&=& {1 \over \sqrt{\pi}} \sin(k_b x + \delta) \quad  x > c + Nd \ .
\label{eq:t18}
\end{eqnarray}
where $E - V_b = \hbar^2 k_b^2/(2 m m^*_b)$ measures the energy above 
the top of the barrier in the asymptotic zone, and $k_w^2 = 2m m^*_w 
E /\hbar^2 $ is the wave number inside the central well. Using the 
transfer matrix $W(d)$ to cross $N$ cells gives 
 \begin{eqnarray}
&&{1\over {\sqrt{\pi}}} \pmatrix{ \sin (k_b(c + Nd) + \delta ) \cr \nu_b \cos 
(k_b(c+Nd)+ \delta)} =   \nonumber \\ 
&& \hskip 2cm B_0 W^{(N)} \pmatrix{\sin k_wc \cr \nu_w\cos k_wc} \ , 
\label{eq:t19}
\end{eqnarray}
where $\nu_w = k_w/ m_w^*$, $\nu_b = k_b/m_b$ and  $W^{(N)} = W^N(d) 
$, so that: 
 \begin{eqnarray}
&&{1\over \nu_b}\tan(k_b(c+Nd) + \delta) = \nonumber \\
&& \hskip 1cm {{W^{(N)}_{11} \sin k_wc + 
W^{(N)}_{12} \nu_w \cos k_wc}\over {W^{(N)}_{21} \sin k_wc + W^{(N)}_{22} 
\nu_w \cos k_wc}} 
 \label{eq:t20}
\end{eqnarray}
determines $\delta$ and
 \begin{eqnarray}
&&{1\over{\sqrt{\pi}}}\sin (k_b(c+Nd) + \delta) = \nonumber \\ 
&& \hskip 2cm B_0 (W^{(N)}_{11} \sin k_wc +W^{(N)}_{12} \nu_w \cos k_wc)
\label{eq:t21}
\end{eqnarray}
gives the normalization $B_0$. Note that the matrix elements of 
$W^{(N)}$ can be easily computed from those of $W(d)$ using 
eq. \ref{eq:t7} (or eq. \ref{eq:t14} 
when the Bloch phase $\phi$ is real). One need not 
solve explicitly for the phase shift $\delta(E)$ because only $|B_0|^2$ is 
required to compute the transition rate, and the identity
$\sin^2 z = \tan^2 z/( 1 + \tan^2 z)$ can be used in eq. \ref{eq:t21}.

With the above equations we can construct the wavefunction 
$\Psi_f(x)$ as follows. Wave functions $v(x),\ w(x)$ in a unit cell 
of the lattice are defined in eq. \ref{eq:t3}. Within the $r$'th cell 
following $x=c$, ($r = 1, 2, ...$) the wave function $\Psi_f(x)$ is 
 \begin{equation}
\Psi_r(x) = A_r v(x-c -rd +d) +  B_r w(x-c -rd +d) 
\label{eq:t22}
\end{equation}
and from the matching at $x=c$ we have: $A_1 = B_0 \sin k_w c$, $B_1 = 
B_0 \nu_w/\nu_b \cos k_w c$. 
In general,  
 \begin{equation}
\pmatrix{A_{r+1} \cr B_{r+1}/m^*} = W(d) \pmatrix{A_{r} \cr 
B_{r}/m^*} \ , \quad r= 1, ... (N-1) \, . 
\label{eq:t23}
\end{equation}
This gives values of  $\Psi_f \to \Psi_r(x)$ in each cell and allows 
the calculation of the matrix element in eq. \ref{eq:t17}. 

The ground state wave function is computed in a similar manner. 
Inside the central well it is
 \begin{eqnarray}
\Psi_0(x) &=& N_0 \cos (k_w x) \quad , \quad |x| < c %\nonumber \\ 
\label{eq:t24}
\end{eqnarray}
and at $x = c + Nd$ it must match to a decaying exponential, 
as in eq. \ref{eq:t6}. If the 
state is well bound it is a good approximation to use the Kalotas 
state which vanishes at the edge of the lattice or the Weber state 
that in principle extends to infinity. The normalization constant 
$N_0$ must be computed by summing the normalization integrals from 
every cell as well as from the central well. If one integrates 
$<v|v>, \, <v|w>, $ and $<w|w>$ over the unit cell, then it is just a 
matter of multiplying these integrals by the coefficients in the 
$r$'th cell and summing. 

Because the effective mass depends both on position and momentum, it 
is not obvious how to evaluate the matrix elements of the transition 
operator $S$. It is reasonable, in the term $p/m^*$ to let the $p$ 
act on the excited state $\psi_E$ and interpret the effective mass as 
being at that energy. Conversely, in the $(1/m^*) p$ term, where the $p$ 
acts on the intial state, we use the ground state effective mass 
$m_0^* $. Then, 
 \begin{eqnarray}
<\psi_E| S |\psi_0> &=& 
- \int_a^b  \frac{d\psi_E}{dx} \frac{1}{m^*(E)} \psi_0  dx + 
\nonumber \\ && 
\int_a^b \psi_E \frac{1}{m^*(E_0)} \frac{d\psi_0}{dx}   dx 
\label{eq:t25}
\end{eqnarray}
Since $\psi'/m^*$ is continuous at interfaces between wells and 
barriers, the integrand is continuous, despite the jumps in $m^*$.  
When the effective mass $m^*$ is piece-wise constant, we can evaluate the 
integral over a series of intervals of constant $m^*$ (here 
interpreted as the $m^*(x,E)$).

The squared matrix element, including the density of states factor, 
is plotted as a function of energy in Fig. \ref{fig10}, which is to 
be compared with Fig. 2 of Sirtori et al. \cite{Cap92}.  (However, 
their figure has normalized the peak intensity to unity in each case, 
obscuring the fact that the integrated strength is constant.) As the 
number of side wells increases, the computed excitation 
function rapidly becomes very narrow. (The experimental width is much 
larger than theory, indicating that something else is happening.) It 
shows that even a small number of cells is sufficient to give a well 
confined state. We also find increasing strength in the second 
allowed band near 700 meV as cells are added. The integrated strength 
under the main peak (from 500 to 640 meV) varies only a few percent. 

To illustrate how the continuum wave functions evolve in the region 
of the CBS, we show in Fig. \ref {fig12} four cases spanning the 
energy range. It can be seen that as one passes over the CBS energy 
at $563$ meV, an additional node appears in the wave function. Away 
from this resonance, (panels (a), (b)) the wave function consists 
mainly of the growing solution in the lattice, so the amplitude is 
largest at the outside edge (where it is fixed, according to eq. 
\ref{eq:t18}). Close to the CBS, (panel (c)), there is a large 
component of the decaying solution, making the amplitude in the 
central well large. Increasing the energy again, (d), brings back 
more of the growing solution. At $577$ meV, the position of the ABS, 
only the growing solution would contribute. If we had more than $N=3$ 
cells, the effects would be even more pronounced.

While the peak in the transition strength becomes very narrow as the 
number of lattice cells $N$ increases from zero to 3, the integrated 
strength is almost constant. This can be understood from the sum rule which 
follows from eq. \ref{eq:t17}, and is discussed in the appendix.  
The total strength, $M_2$, is defined in eq. \ref{eq:a2}.
In addition to the integral over the continuum, when there are $N$ 
Bragg reflectors on either  side of the central defect, there will be 
$N$ discrete odd-parity bound states, which must also be included in 
the sum. Typically these account for something like $6 \%$ of the total 
strength. These odd bound states are shown in Fig. \ref{fig11}, for 
the case $N=2$.  In this figure, the wave functions are remarkably similar 
inside the region where the ground state is large, so they give almost 
equal contributions to the sum rule. 

Turning now to the results, one has to distinguish between the 
no-reflector case and the $N$-reflector case. In the former, the 
integrated transition strength (ITS), $M_2$, is about $1.4 \AA^{-
2}$. The strength is very broadly distributed above threshold; see 
line (a) in Fig. \ref{fig10}. We have looked at $N = 1$ to $5$ cells 
on each side. %    (for example, see columns 2, 5, and 8 of Table \ref{tx6}). 
For these cases, the ITS is around $1.35 \AA^{-2}$, of 
which about $0.09$ comes from the bound odd-parity excited states. As 
$N$ increases, the ITS fluctuates only a little. 
The strength remains highly concentrated into the CBS peak (about $80 
\%$), the remainder being spread quite widely ($6 \%$ in the bound 
states and $14 \%$ in the continuum). 

Because the  CBS peak becomes so narrow, we estimate the integral 
under it by assuming a Breit-Wigner shape, and deducing the height 
and width from the calculations. 

The sum rule, calculated according to eq. \ref{eq:a3}, is always 
about 6 to 10 \% higher than the ITS, if we (arbitrarily) set the 
doorway state mass $m_E$ to be at $E = E_{CBS}= 563$ meV. 
The main term in eq. \ref{eq:a3} is proportional to 
$(1/m_0 + 1/m_E)^2$, so we can easily adjust the doorway energy to 
ensure that the sum rule will agree with the ITS. We call this the 
effective doorway energy $E_D$. With no reflectors $E_D$ is about 740 
meV (535 above the ground state energy). This is reasonable since the 
excitation strength is very broadly distributed above threshold. With 
one reflector (on each side), $E_D$ drops to 660 meV; but then it 
slowly rises, at least up to $N = 5$, where it reaches 670 meV. 

\section{Optimal CBS confinement } 

In this section we discuss some general principles for designing a 
well confined CBS. Consider a general unit cell of width $d$, 
within which the potential and the effective mass are arbitrary 
functions of $x$. (Since we will fix the CBS energy, this allows for 
energy dependence of the effective mass.) Now let us arbitrarily 
divide the cell into two parts so that the widths $a$ and $b$ add to 
$d$, and denote by $W^a$, $W^b$ the transfer matrices of the two 
parts. When $a$ is on the left, we have $W^d = W^b \, W^a$, with 
elements 
 \begin{eqnarray}
W^d_{i\, j} &=& W^b_{i\, 1} W^a_{1\, j} + W^b_{i\, 2}  W^a_{2\, j}  
\label{eq:t26}
\end{eqnarray}
If the whole array is symmetric about the origin, there will be 
two type-$a$ portions together at the origin, and the sequence of 
potentials is $... baba ... ba|ab ... abab ...$. When we look for odd 
parity confined states of such an array, it is equivalent to putting 
a hard wall at the origin, and solving only the right side. The wave 
function at the edge of the first cell will be, using eq. \ref{eq:t3} 
 \begin{eqnarray}
\pmatrix{\psi \cr \psi^{'}/m^*}\Bigr|_d &=& \pmatrix{W^d_{12} \cr W^d_{22} } 
= \lambda \pmatrix{0 \cr 1} 
\label{eq:t27}
\end{eqnarray}
The second equality holds if we imagine an infinite array, and demand 
an eigenstate with the wave function in each cell 
differing only by a factor $\lambda$. This wave function will vanish 
at both $x = 0$ and $d$, and in the second cell, the value of 
$\psi'/m^*$ will differ by a factor $\lambda$ from the first. 
The condition that must be satisfied is $W^d_{12} = 0$, and then 
$\lambda = W^d_{22}$. Writing these out in terms of the component 
transfer matrices gives 
 \begin{eqnarray}
W^d_{1\, 2} &=& W^b_{1\, 1} W^a_{1\, 2} + W^b_{1\, 2}  W^a_{2\, 2}  = 
0 \quad {\rm or } \nonumber \\ 
\frac{W^b_{1\, 1}}{W^b_{1\, 2}} &=&  - \frac{W^a_{2\, 2}}{W^a_{1\, 2}} 
\quad {\rm and } \nonumber \\ 
W^d_{2\, 2} &=& W^b_{2\, 1} W^a_{1\, 2} + W^b_{2\, 2}  W^a_{2\, 2} = 
\lambda  \nonumber \\ 
&=& \left( W^b_{2\, 1}  + W^b_{2\, 2}  \frac{W^a_{2\, 
2}}{W^a_{1\, 2}}\right)  W^a_{1\, 2}  \nonumber \\ 
&=& \left( W^b_{2\, 1}  -  W^b_{2\, 2}  \frac{W^b_{1\, 1}}{W^b_{1\, 2}}
\right)  W^a_{1\, 2}  \nonumber \\ 
&=& \left( W^b_{2\, 1}W^b_{1\, 2}  -  W^b_{2\, 2} W^b_{1\, 1} 
\right)  \frac{W^a_{1\, 2}}{W^b_{1\, 2}}  \nonumber \\ 
\lambda &=& - \frac{W^a_{1\, 2}}{W^b_{1\, 2}} = -\frac{w_a(a)}{w_b(b)} \quad . 
\label{eq:t28}
\end{eqnarray}
In the last step we have used the form of $W$ as in eq. \ref{eq:t3}. 

What this tells us is that to make the eigenvalue as small as 
possible, we must make the $w_a(x)$ solution as small as possible at 
the right edge of the $a$ part-cell, and conversely, $w_b(x)$ as 
large as possible at the right edge of the $b$ part-cell. 

The above is true for {\it any} division of the cell into two parts. 
In the system studied by Capasso et al., the logical division is into 
the two layers of GaInAs and AlInAs. In that situation, the 
off-diagonal elements have the form $W^c_{1, 2} = - \sin k_c 
c/\nu_c$ where $\nu_c = k_c/m^*_c$ ($c = a,\, b$) is the velocity. If the 
Bragg reflection condition holds, then $\sin k c = \pm 1$ and 
$\lambda = - \nu_b/\nu_a$ is just the ratio of velocities in the 
two parts of the cell. (This is analogous  to the problem of waves on 
a string, with one part thin and the other thick. At the join, the 
displacement $y(x)$ is continuous, and the ratio of the slopes 
$y'(L)/y'(R)$ is the ratio of the velocities squared.) 
One sees that in the $a$-cell, the solution $w_a(x)$ rises to the 
value $w_a(a) = 1/\nu_a$, while starting from $d$ and moving backwards 
through the barrier region, the corresponding solution falls to the 
value $w_b(-b) = - w_b(b) = -1/\nu_b$. Normalizing the $b$ solution 
to ensure continuity at $x = a$ requires the factor $\lambda$.  

The result eq. \ref{eq:t28} is quite amazing because in the general 
situation where the potential and effective mass vary arbitrarily, 
the dividing line can be placed wherever you wish. To make the 
eigenvalue small, one must make the ``odd-parity" solution $w(x)$ as 
large as possible everywhere in the second part-cell and as small 
as possible in the first. As observed by Weber, the first aspect 
can be achieved by choosing an energy just above the barrier 
(small $k_b$). To meet the Bragg condition, this forces a large $b$, 
and the linear variation of $w_b(x)$ over the barrier leads to a 
large wave function at $x=b$. That is why lowering the energy of the 
CBS in general improves confinement. 

However, our aim is to improve confinement while {\it keeping the 
energy} of the CBS {\it fixed}. To do so, we are guided by Shockley's 
explanation for the existence of surface states, as discussed in 
\cite{SSWM00}. We split the $a$ part-cell into two sections, $a_1$, 
$a_2$, making the left side more attractive, and the right side less 
so ($V_{w1} < V_w < V_{w2}$). Then the greater curvature of $w_a$ 
near the origin, balanced by less curvature to the right, will lead 
to a smaller value of $w_a(a)$, even if the average attraction is the 
same. We leave the $b$-cell fixed, but a similar strategy with less 
repulsion on the right side, can obviously be employed there.

To illustrate this, we have selected a set of 
heterostructures based on quaternary alloys $Ga_x In_{1-x} As_y P_{1-
y}$, lattice-matched to $InP$ ($x = 0.468 y$). 
We took information from Figs. 1.17 (for band gaps) and 1.20 (for 
effective masses) of Swaminathan and Macrander \cite{SM91}. 
For the band offsets, S. Adachi \cite{SA92} gives 
 \begin{eqnarray}
\Delta E_c &=& 268 y + 3 y^2 \nonumber \\ 
\Delta E_v &=& 502 y - 152 y^2 \quad . 
\label{eq:t29}
\end{eqnarray}
The band alignments are also discussed on p. 87 of Davies \cite{JHD98}. 
Putting together this information, we arrived at the following set of 
parameters: 
 \begin{eqnarray}
y = 1.0 && \quad m^*_0 = 0.043 \quad \bar{E}_g =0.880\,\,{\rm eV} \nonumber \\ 
y = 0.5 && \quad m^*_0 = 0.061 \quad \bar{E}_g =1.080\,\,{\rm eV} \nonumber \\ 
y = 0.0 && \quad m^*_0 = 0.081 \quad \bar{E}_g =1.360 \,\,{\rm eV} 
\label{eq:t30}
\end{eqnarray}
where $\bar{E}_g$ are effective band gaps in the sense of 
Nelson et al. \cite{NMK87}. 

In this way we can have conduction band potential steps of 125 meV 
(from $Ga_{.47} In_{.53} As$ to $Ga_{.23} In_{.77} As_{.5} P_{.5}$ 
%with $y = 0.5$, $m^* = 0.061, \bar{E_g} = 1080$) 
or 250 meV to $In P$ ($y = 0$).  % with $m^* = 0.081, \bar{E_g} = 1380$. 
These are a quarter (denoted $Q$) or a half (denoted $H$) step up to 
the $500$ meV barrier of $Al_{.48} In_{.52} As$. 

As our baseline, (denoted $(Q,Q)$ below), we take the $a$-well to 
consist entirely of $Q$ (y= 0.5) material, so the potential floor at 
$125$ meV is $375$ meV below the barrier. The barrier width was fixed 
at $44.3 \AA$ which satisfies the Bragg condition. For a width $a = 
15.82 \AA$, the CBS is at 63 meV above the top of the barrier, as in 
the original experiment. The eigenvalue $\lambda = -0.4026$, which is 
not as favorable as in the original work because the well is not so 
deep. (Capasso et al. selected the materials to have the greatest 
possible well-barrier potential difference.) The potential properties 
are summarized in Table \ref{tx5}, top line. 

Next we divide the $a$-well into two parts, one of $GaInAs$ ($y=1$) 
and the other of $InP$. We adjusted the widths $a_1$ and $a_2$ to 
keep $E_{CBS}$ fixed. In the second line of Table \ref{tx5}, 
denoted $(0,H)_1$, the two part wells are of equal width $9.353 \AA$, 
and the eigenvalue is $\lambda = -0.396$.  In the third line, denoted 
$(0,H)_2$, the deeper well has width $a_1 = 10.50 \AA$, and the 
shallower part $a_2 = 7.911 \AA$, giving $\lambda = -0.374$. This may 
seem a small gain, but we shall see that the improvement is 
significant.  

We then computed the CBS properties for a finite array based on the 
above materials, with results shown in Tables \ref{tx6}, \ref{tx7} 
and Fig. \ref{fig13}. The transition strength is significantly 
narrower and more strongly peaked for the split-well examples. The 
strength to the bound states in the split-well cases was only 25-30\% 
of that of the (Q,Q) reference case, and the strength to the 
continuum states was much larger.  The results are summarized 
in Table \ref{tx6}, where the number of cells means the number of 
Bragg reflectors placed on each side of the central defect. 

The differences in the total strength are reflected in the portion 
concentrated in the CBS peak.  The decay constant strongly 
influences both the sharpness of, and the area under, the 
peak.  The cases with a lower (in magnitude) decay constant have 
narrower, and larger (in terms of area) peaks in the transition 
strength curve as seen in Table \ref{tx7}.

Overall, the decay constant has a significant effect on 
both the total strength, and its continuum and bound state contributions 
individually.  A lower value of the decay constant (in magnitude) 
results in better confinement of the CBS as is evidenced by the width 
of the peak in the transition strength curve.  We conclude that the 
split-well strategy can produce much better confinement of the CBS. 
It should be feasible to confirm this method of improving the 
confinement of CBS, experimentally.

\section{Conclusion}

We have shown that Continuum Bound States are closely related to 
surface states, because both arise as a result of perturbing an 
infinite periodic system. The results of Weber\cite{Weber94} 
concerning the experiment of Capasso et al. were verified. However, 
while we confirm that a CBS can exist even when the Bragg conditions 
are not well satisfied, we note that for this type of potential, 
one can prove analytically that optimal confinement is 
achieved by the Bragg conditions. In addition, by enclosing a finite 
array in a box, we have traced the evolution of the Bloch states in 
the allowed bands as the number of Bragg layers is increased. 

We have derived a sum rule, within the conduction band only 
model, that explains the integrated transition strength from the 
ground state to the continuum in the region where the CBS exists. 
About 70\% of the sum rule strength is concentrated in this region.  
Finally, we have identified the factors that allow one to improve the 
confinement of a continuum bound state, and proposed a way of testing 
this. 

%\vskip 0.5 cm 

\acknowledgements

We are grateful to NSERC-Canada for research 
grant SAPIN-3198 (DWLS) and Summer Research Awards (PJ, JS), and to 
DGES-Spain for continued support through grant
PB97-0915 (JM). This work was carried out as part of 
CERION, an Esprit project EP-27119 funded by the EU and NSERC. 

\appendix

\section{Sum Rule for intra-subband transitions}

Sirtori et al. \cite{SCFS94} 
discussed the sum rule for excitations to the CBS within a two-band 
Kane model. Here we limit our discussion to what can be done within a 
conduction-band only model, eq. \ref{eq:t1}.  The difficulty which 
arises in this context is that having an energy-dependent effective mass 
means that the Hamiltonian does not have a complete orthonormal set 
of excited states, so the sum rule can only be approximate.

We take the operator to be 
 \begin{eqnarray}
-i\hbar S &=& (p \frac{1}{m_E} + \frac{1}{m_0} p)
\label{eq:a1}
\end{eqnarray} 
(cf. eq. \ref{eq:t16}.) Then the sum rule is 
 \begin{eqnarray}
&&\hbar^2 M_2 = \hbar^2 <0| S^\dagger S |0> \nonumber \\ 
 &=& <0| (p \frac{1}{m_0} + \frac{1}{m_E} p) (p \frac{1}{m_E} + \frac{1}{m_0} 
p ) |0> \nonumber \\ 
&=& <0| p \frac{1}{m_0} \frac{1}{m_0} p|0> + 
    <0| p \frac{1}{m_0} \frac{1}{m_E} p|0> - 
\nonumber \\ 
&&    i\hbar <0| p \frac{1}{m_0} (\frac{1}{m_E}))^{'}|0> + 
     <0| p \frac{1}{m_E} \frac{1}{m_0} p|0> + \nonumber \\ 
&& 
   i\hbar <0| (\frac{1}{m_E})^{'} \frac{1}{m_0} p|0> + 
<0| p \frac{1}{m_E} \frac{1}{m_E} p|0> + \nonumber \\ 
&&
   i\hbar <0| (\frac{1}{m_E})^{'} \frac{1}{m_E} p|0> - 
   i\hbar <0|p \frac{1}{m_E}  (\frac{1}{m_E})^{'} |0> 
\nonumber \\ 
&&  +  \hbar^2 <0|(\frac{1}{m_E})^{'}  (\frac{1}{m_E})^{'} |0>  \, ,
\label{eq:a2}
\end{eqnarray} 
\begin{eqnarray} 
M_2 &=& \int |\psi_0^{'}|^2 (\frac{1}{m_0} + \frac{1}{m_E} )^2 dx + 
  <0|(\frac{1}{m_E})^{'}  (\frac{1}{m_E})^{'} |0>  \nonumber \\ 
&& + 2 \int \psi_0 \psi_0^{'} (\frac{1}{m_0} + \frac{1}{m_E})  
(\frac{1}{m_E})^{'}\, dx  \quad . 
\label{eq:a3}
\end{eqnarray} 
To obtain this expression we moved the $p$-operators 
until they act on the ground state wave function directly. In the 
case of a constant effective mass, $m_0 = m_E = 1$, only the first 
integral survives. In this case the sum rule must be exact, and we 
found close agreement between the sum rule expression, 
and direct integration: 
 \begin{eqnarray}
M_2 &=& \int <\psi_0| S^\dagger |\psi_E> \frac{dk}{dE} <\psi_E| 
S |\psi_0> dE    \nonumber \\ 
{\rm where} && \nonumber \\ 
\frac{d k}{dE} &=&  \frac{1}{2k} \frac{2m}{\hbar^2} m_{0b} 
\bigl( 1 + \frac{2(E - V_b)}{E_b} \bigr)
\label{eq:a4}
\end{eqnarray} 
which includes non-parabolicity. The wave number $k$ is defined by the 
energy above the barrier. At high energy the density of states tends 
to a constant, rather than going to zero, as it would for constant 
mass. The case without energy-dependence can be recovered if $E_b 
\to \infty$. The density of states factor was tested by doing the 
integrals either over energy or over wave number $k$.

When we introduce $x$-dependence to the effective mass, the terms 
involving the derivative of $(1/m^*_E)$ contribute. Because in this 
model the mass is piecewise constant, the derivative is a Dirac delta 
function times the discontinuity in $(1/m^*_E)$. The integral 
in the last line of eq. \ref{eq:a3} is then a sum of values 
evaluated at the layer edges. 

When we introduce energy-dependence as well, both here, and in the 
first integral, factors such as $\psi_0^{'}/m_E$ are discontinuous, 
because the mass $m_E$ is taken at one energy and the ground state 
wave function at another. To resolve this ambiguity we took the 
average of the two values on either side of the discontinuity. For 
these materials, the well and barrier masses are similar, so it is 
not a large uncertainty. This is the stage at which the sum rule can 
only be approximate. Moreover, we need a prescription for the energy 
$E$ at which we evaluate $m_E$.  Thinking in terms of the doorway 
state approximation, initially we took the CBS energy.  

The second term in eq. \ref{eq:a3} involves the square of a Dirac 
delta function, and is undefined. We simply omit this contribution.

%\vskip 1cm

\begin{table}[htbp]
%\begin{center}
\caption{CBS properties varying widths $a$ and 
$b$ separately, for a central well.} 
\begin{tabular}{|c|c|c|c|c|c|} 
\multicolumn{6}{|c|} {Central Well} \\ \hline
a & b (\AA) & E (meV) & $\lambda_{-}$ & $k_w a$ & $k_b b$ 
\\ \hline %\hline
16 & 16  & 686 & -0.565 & 1.876 & 1.011 \\ %\hline
16 & 20  & 645 & -0.499 & 1.797 & 1.105 \\ %\hline
16 & 30  & 589 & -0.389 & 1.684 & 1.275 \\ %\hline
16 & 39  & 563 & -0.327 & 1.632 & 1.383 \\ %\hline
16 & 50  & 545 & -0.277 & 1.595 & 1.483 \\ %\hline
16 & 60  & 534 & -0.244 & 1.574 & 1.557 \\ %\hline
16 & 75  & 525 & -0.210 & 1.555 & 1.647 \\ %\hline
16 & 100 & 516 & -0.173 & 1.537 & 1.767 \\ %\hline
16 & 125 & 512 & -0.150 & 1.528 & 1.862 \\ %\hline
16 & 150 & 509 & -0.135 & 1.522 & 1.942 \\ %\hline
16 & 200 & 505 & -0.115 & 1.516 & 2.072 \\ \hline %\hline
10 & 39  & 666 & -0.601 & 1.149 & 2.320 \\ %\hline
11 & 39  & 651 & -0.545 & 1.243 & 2.202 \\ %\hline
12 & 39  & 635 & -0.493 & 1.332 & 2.071 \\ %\hline
13 & 39  & 618 & -0.444 & 1.416 & 1.926 \\ %\hline
14 & 39  & 600 & -0.400 & 1.494 & 1.765 \\ %\hline
15 & 39  & 581 & -0.361 & 1.566 & 1.585 \\ %\hline
17 & 39  & 544 & -0.298 & 1.693 & 1.148 \\ %\hline
18 & 39  & 525 & -0.273 & 1.750 & 0.860 \\ %\hline
19 & 39  & 506 & -0.251 & 1.801 & 0.426 \\ \hline
\end{tabular}
%\end{center}
\label{tx1}
\end{table}

\begin{table}[htbp]
%\begin{center}
\caption{CBS properties varying widths $a$ and 
$b$ separately for a central barrier. }
\begin{tabular}{|c|c|c|c|c|c|} 
\multicolumn{6}{|c|}{Central Barrier} \\ \hline
a & b (\AA) & E (meV) & $\lambda_{-}$ & $k_w a$ & $k_b b$ 
\\ \hline %\hline
16 & 16  & 778 & -0.626 & 2.058 & 1.273 \\ %\hline
16 & 20  & 715 & -0.555 & 1.934 & 1.372 \\ %\hline
16 & 30  & 620 & -0.430 & 1.747 & 1.496 \\ %\hline
16 & 39  & 577 & -0.354 & 1.661 & 1.539 \\ %\hline
16 & 50  & 549 & -0.289 & 1.605 & 1.561 \\ %\hline
16 & 60  & 535 & -0.246 & 1.575 & 1.570 \\ %\hline
16 & 75  & 523 & -0.200 & 1.550 & 1.575 \\ %\hline
16 & 100 & 513 & -0.153 & 1.530 & 1.577 \\ %\hline
16 & 125 & 508 & -0.123 & 1.521 & 1.577 \\ %\hline
16 & 150 & 506 & -0.103 & 1.516 & 1.576 \\ %\hline
16 & 200 & 503 & -0.077 & 1.511 & 1.575 \\ \hline %\hline
10 & 39  & 602 & -0.443 & 1.070 & 1.785 \\ %\hline
11 & 39  & 597 & -0.416 & 1.169 & 1.734 \\ %\hline
12 & 39  & 592 & -0.396 & 1.268 & 1.689 \\ %\hline
13 & 39  & 588 & -0.379 & 1.367 & 1.648 \\ %\hline
14 & 39  & 584 & -0.368 & 1.465 & 1.609 \\ %\hline
15 & 39  & 581 & -0.359 & 1.563 & 1.573 \\ %\hline
17 & 39  & 574 & -0.351 & 1.757 & 1.505 \\ %\hline
18 & 39  & 571 & -0.352 & 1.854 & 1.472 \\ %\hline
19 & 39  & 568 & -0.356 & 1.950 & 1.439 \\ \hline
\end{tabular}
%\end{center}
\label{tx2}
\end{table}

\begin{table}[htbp]
%\begin{center}
\caption{CBS properties for fixed energy 560 eV. } 
\begin{tabular}{|c|c|c|c|c|c|} 
\multicolumn{6}{|c|}{Energy fixed at 560 meV} \\ \hline
a & b (\AA) & Energy & $\lambda_{-}$ & $k_w a$ & $k_b b$ \\  
\hline

10     & 77.29  & 560.00 & -0.591 & 1.016  & 2.671 \\ %\hline
13     & 65.14  & 560.00 & -0.393 & 1.321  & 2.251 \\ %\hline
14     & 58.25  & 559.99 & -0.345 & 1.423  & 2.013 \\ %\hline
15     & 49.70  & 559.98 & -0.318 & 1.524  & 1.717 \\ %\hline
15.4   & 45.98  & 560.00 & -0.315 & 1.565  & 1.589 \\ %\hline
15.45  & 45.51  & 560.01 & -0.315 & 1.570  & 1.573 \\ %\hline
15.455 & 45.465 & 560.01 & -0.315 & 1.5706 & 1.5713 \\ %\hline
15.457 & 45.449 & 560.00 & -0.315 & 1.57084& 1.57065 \\ %\hline
15.46  & 45.42  & 560.00 & -0.315 & 1.5712 & 1.5597 \\ %\hline
15.5   & 45.05  & 560.00 & -0.315 & 1.575  & 1.557 \\ %\hline
16     & 40.45  & 559.96 & -0.319 & 1.626  & 1.397 \\ %\hline
17     & 31.99  & 560.01 & -0.348 & 1.727  & 1.106 \\ %\hline
18     & 25.25  & 560.01 & -0.398 & 1.829  & 0.873 \\ %\hline
19     & 20.16  & 560.02 & -0.460 & 1.931  & 0.697 \\ %\hline
20     & 16.33  & 560.03 & -0.527 & 2.033  & 0.564 \\ \hline
\end{tabular}
%\end{center}
\label{tx3}
\end{table}

 \begin{table}[htbp]
\caption{Optimized three-layer potentials indicating the changes in 
the widths $a_1,\ a_2$, $w(a)$, $\nu_a = k_a/m^*_a$, and 
eigenvalue $\lambda$. In all cases $b = 44.2851450 \AA$, $w_b(b) = 
2.1450935 $.\ and $E_{CBS} = 563.0$ meV.} 
 %\begin{center}
\begin{tabular}{|r|c|c|c|c|c|}
%\hline
 case  & $a_1$  & $a_2$    &   $w(a)$   &  $\nu_a$  &   $\lambda$ \\ 
\hline
 $(Q,Q)$  & 7.9109 & 7.9109 & 0.8636  & 1.1579 & -0.4026 \\
 $(0,H)_1$& 9.3535 & 9.3535 & 0.8495  & 0.9092 & -0.3960 \\
 $(0,H)_2$& 10.4995 & 7.9109 & 0.8031 & 0.9092 & -0.3744 \\
%\hline
 \end{tabular}
%\end{center}
\label{tx5}
\end{table}

\begin{onecolumn}

\begin{table}[htbp]
\caption{Evolution of the transition strength ($\AA^{-2}$) with increasing 
number of Bragg reflectors.}
\begin{center}
\begin{tabular}{|c|c|c|c|c|c|c|c|c|c|} %\hline 
\multicolumn{1}{|c|}{} & 
\multicolumn{3}{c|}{Bound} & 
\multicolumn{3}{c|}{Continuum} & 
\multicolumn{3}{c|}{Total Strength} \\ \hline 
\# Cells & (Q,Q) & $(0,H)_1$ & $(0,H)_2$ & (Q,Q) & $(0,H)_1$ & $(0,H)_2$ & (Q,Q)
 & $(0,H)_1$ & $(0,H)_2$ \\ \hline %\hline
0 & 0 & 0 & 0 & 0.912 & 1.279 & 1.324 & 0.912 & 1.279 & 1.324 \\ %\hline
1 & 0.108 & 0.0267 & 0.0319 & 0.762 & 1.188 & 1.234 & 0.870 & 1.215 & 
1.266 \\ %\hline
2 & 0.117 & 0.0281 & 0.0334 & 0.759 & 1.188 & 1.233 & 0.876 & 1.216
& 1.267 \\ %\hline
3 & 0.118 & 0.0282 & 0.0335 & 0.754 & 1.183 & 1.229 & 0.872 & 1.211 
& 1.263 \\ %\hline
4 & 0.118 & 0.0282 & 0.0335 & 0.746 & 1.176 & 1.224 & 0.864 & 1.204 
& 1.257 \\ %\hline
5 & 0.118 & 0.0282 & 0.0335 & 0.745 & 1.174 & 1.225 & 0.863 & 1.202 
& 1.258 \\ %\hline
\end{tabular} 
\end{center}
\label{tx6}
\end{table}

\begin{table}[htbp]
\caption{Total strength under the CBS peak: dependence on number of Bragg 
reflectors.}
\begin{center}
\begin{tabular}{|c|c|c|c|c|c|c|c|c|c|} %\hline 
\multicolumn{1}{|c|}{}&
\multicolumn{3}{c|}{Peak Height\,\, eV$^{-1} \AA^{-2}$}&
\multicolumn{3}{c|}{Width $\Gamma$ (eV)}&
\multicolumn{3}{c|}{Peak/Total Strength (\%)} \\ \hline 
Cells & (Q,Q) & $(0,H)_1$ & $(0,H)_2$ & (Q,Q) & $(0,H)_1$ & $(0,H)_2$ & (Q,Q) & 
$(0,H)_1$ & $(0,H)_2$\\ \hline %\hline
1 & 27.5 & 38.2 & 45.6 & $1.73*10^{-2}$ & $1.49*10^{-2}$& $1.33*10^{-2}
$ & 85.8 & 73.7 & 75.5 \\ %\hline 
2 & 165. & 247. & 328. & $2.42*10^{-3}$ & $2.07*10^{-3}$ & $1.68*10^{-3}
$ & 71.6 & 66.1 & 68.4 \\ %\hline 
3 & 1026. & 1609. & 2383. & $3.81*10^{-4}$ & $3.16*10^{-4}$ & $2.33*10^{-4}
$ & 70.5 & 66.0 & 69.1 \\ %\hline 
4 & 6300 & 10240 & 16970. & $6.18*10^{-5}$ & $4.92*10^{-5}$ & $3.24*10^{-5}
$ & 70.8 & 65.7 & 68.7 \\ %\hline 
5 & 38950 & 65450 & 121300 & $1.00*10^{-5}$ & $7.76*10^{-6}$ & $4.54*10^{-6}
$ & 71.0 & 66.4 & 68.8 \\ %\hline  
 \end{tabular} 
\end{center}
\label{tx7} 
\end{table}

\end{onecolumn}

\newpage
%\vskip 1.5cm
%\centerline{\bf Figure captions}
\begin{onecolumn}

\begin{figure}[htb]
\begin{center}
\leavevmode
\epsfxsize=8cm
\epsffile{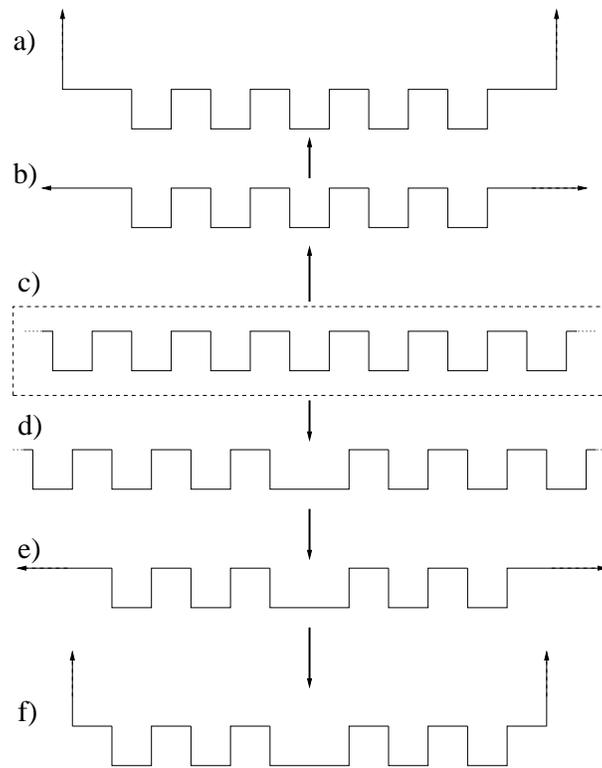}
\caption{Schematic drawing of an infinite array (line c), truncated 
to a finite array (b) and enclosed in walls (a); or with a defect 
(d), also truncated (e) and enclosed (f). } 
 \label{fig1} 
\end{center}
\end{figure}

\begin{figure}[htb]
\begin{center}
\leavevmode
\epsfxsize=8cm
\begin{tabular}{|cc|} \hline
A) & \epsffile{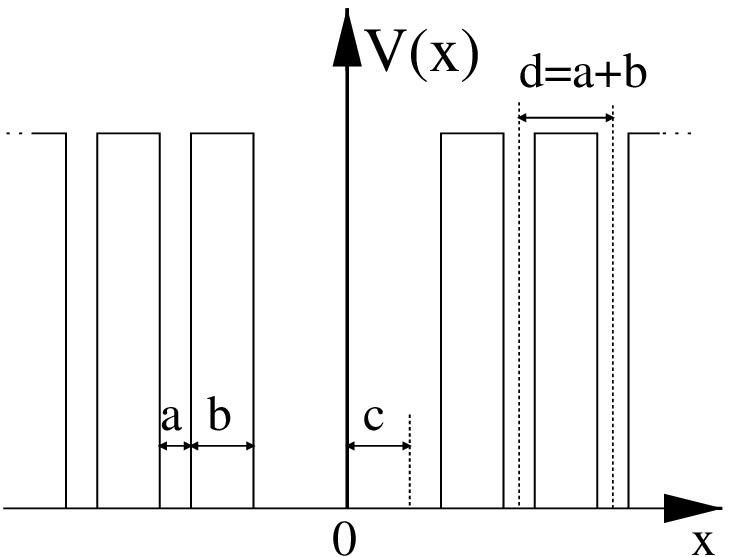} \\ \hline
B) & \epsffile{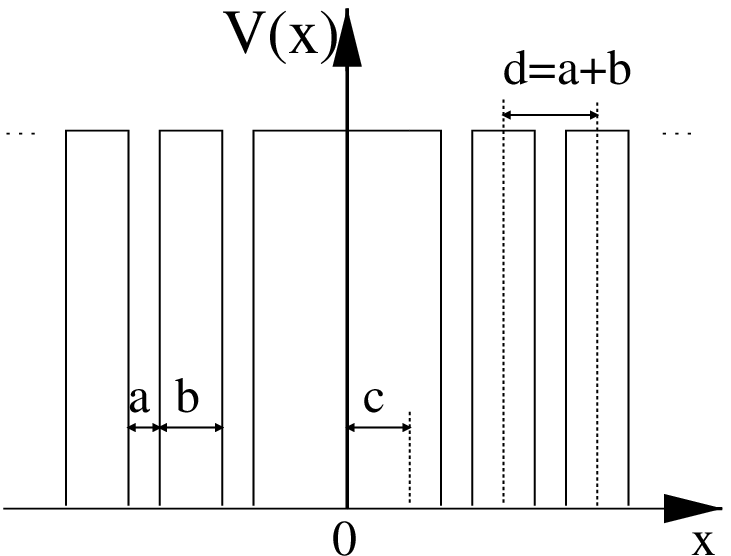} \\ \hline
%B) & \includegraphics[bb=0 0 3in 2.3in,scale=1]{arrayb.eps} \\ \hline
\end{tabular}
\end{center}
\caption{A unit cell of the infinite lattice for A) a well as the central
defect, and B) a barrier as the central defect.}
 \label{fig2} 
\end{figure}

\begin{figure}[htb]
\begin{center}
\leavevmode
\epsfxsize=8cm
\epsffile{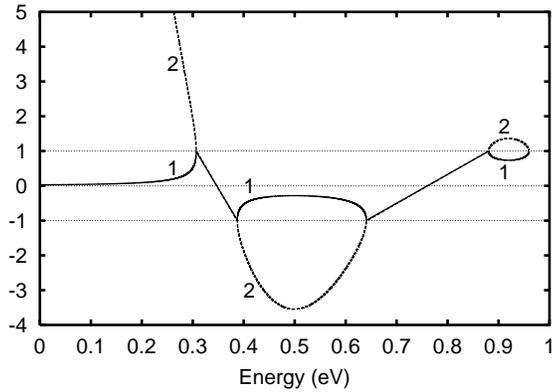}
%\fbox{\includegraphics[bb=0 0 8in 5in,scale=0.5]{evalues.eps}}
\end{center}
\caption{Eigenvalues vs. energy for a central well, $c = 8 \AA$,
showing the smaller magnitude eigenvalue $\lambda_{-}$ (line 1) and 
the larger $\lambda_{+}$ (line 2). The straight lines across the 
allowed zones (where the $\lambda_\pm$ are complex) connect related 
solutions.} 
 \label{fig3} 
\end{figure}

\begin{figure}[htb]
\begin{center}
\leavevmode
\epsfxsize=8cm
\epsffile{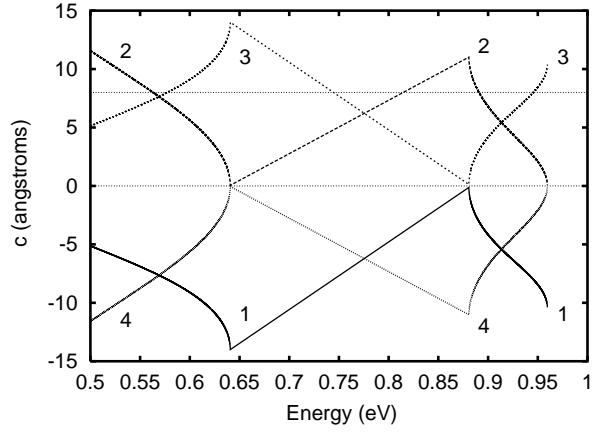}
%\fbox{\includegraphics[bb=0 0 8in 6in,scale=0.5]{cbsfig1.eps}}
\end{center}
\caption{Central well width c versus energy showing the even CBS (line 1), the
odd CBS (line 2), the even ABS (line 3), and the odd ABS (line 4).  See also
figure 6. The straight lines across the allowed zones connect related 
solutions.} 
 \label{fig4} 
\end{figure}

\begin{figure}[htb]
\begin{center}
\leavevmode
\epsfxsize=8cm
\begin{tabular}{|cc|} \hline
a) & \epsffile{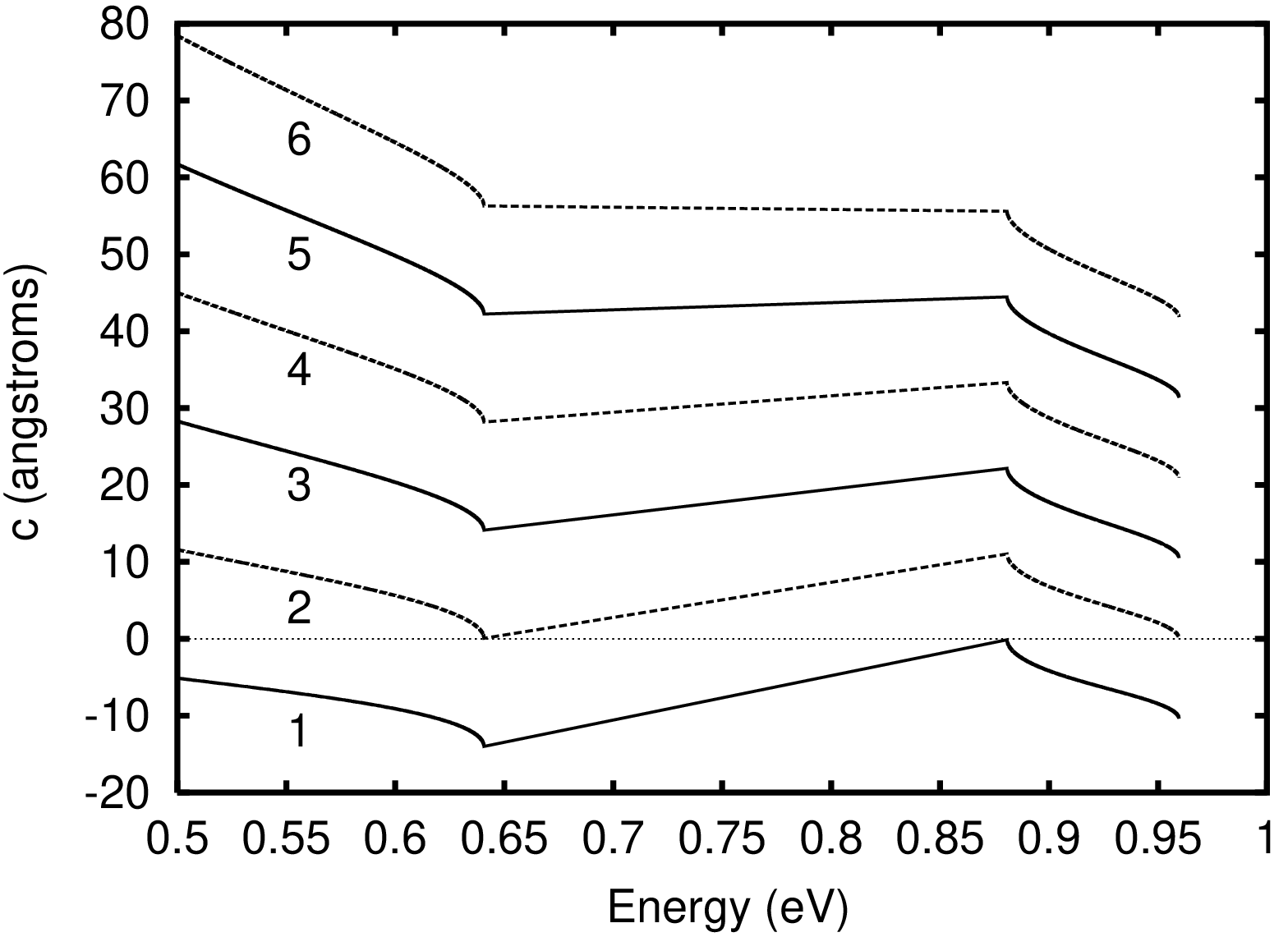} \\ \hline
b) & \epsfxsize=8cm \epsffile{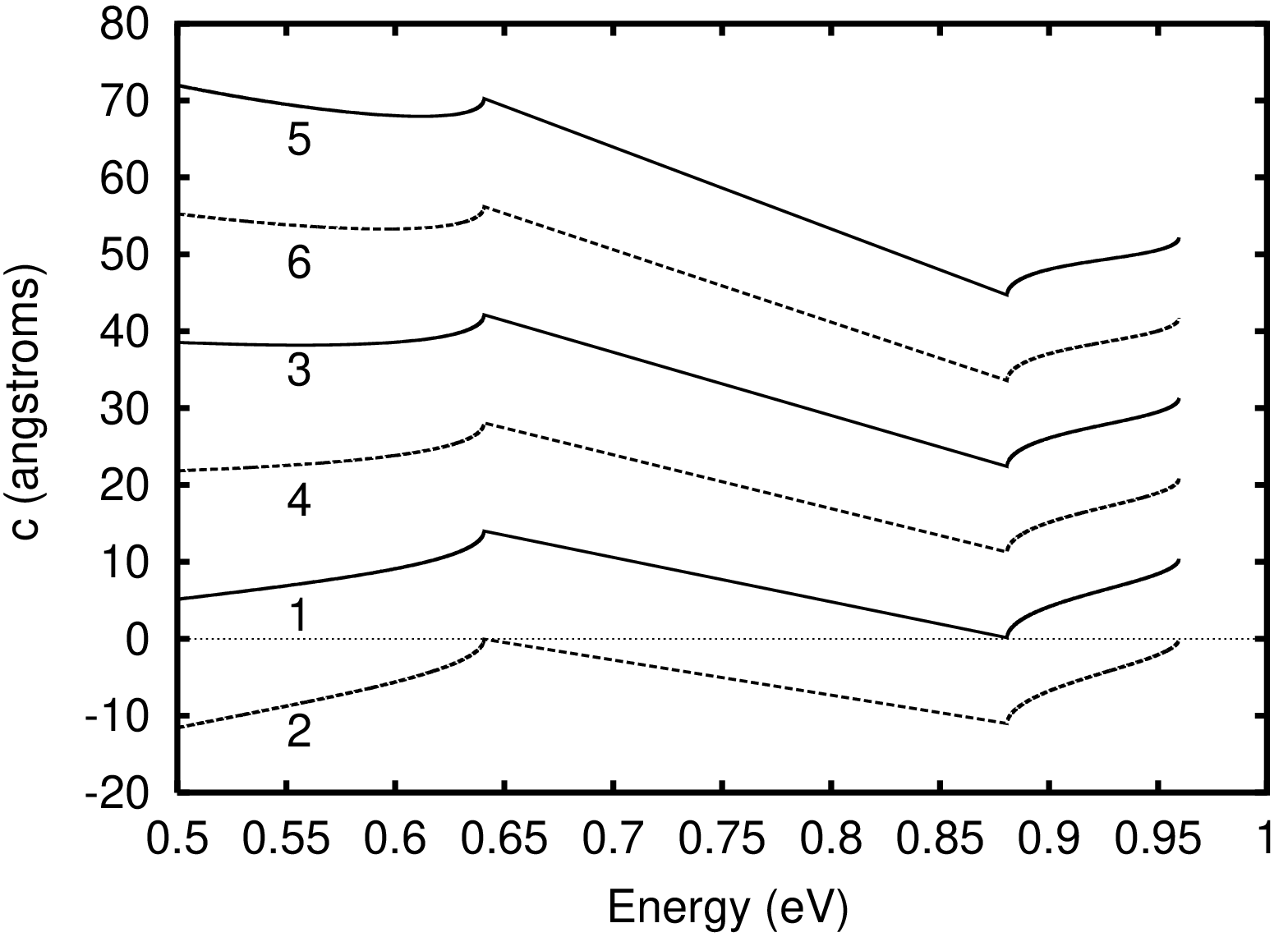} \\ \hline
%a) & \includegraphics[bb= 0 0 8in 6in,scale=0.5]{cbsfig2a.eps} \\ \hline
%b) & \includegraphics[bb=0 0 8in 6in,scale=0.5]{cbsfig2b.eps} \\ \hline
\end{tabular}
\end{center}
\caption{Central well width c versus energy showing a) the CBS and b) 
the ABS.  Odd numbered lines are for even states, and vice versa. The 
straight lines across the allowed zone connect related solutions.} 
\label{fig5} 
\end{figure}

\begin{figure}[htb]
\begin{center}
\leavevmode
\epsfxsize=8cm
\epsffile{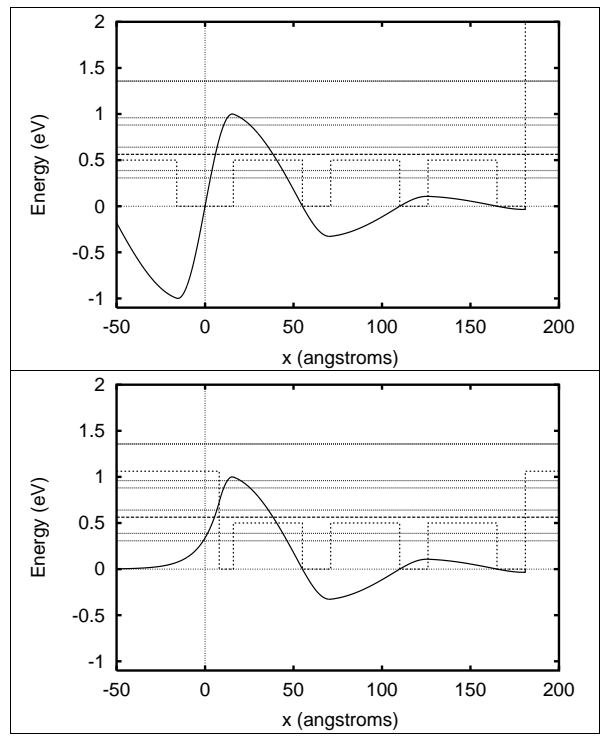}
\end{center}
\caption{Wave functions of a CBS (a) and a surface state (b) compared.}
 \label{fig6} 
\end{figure}

\begin{figure}[htb]
\begin{center}
\leavevmode
\epsfxsize=8cm
\epsffile{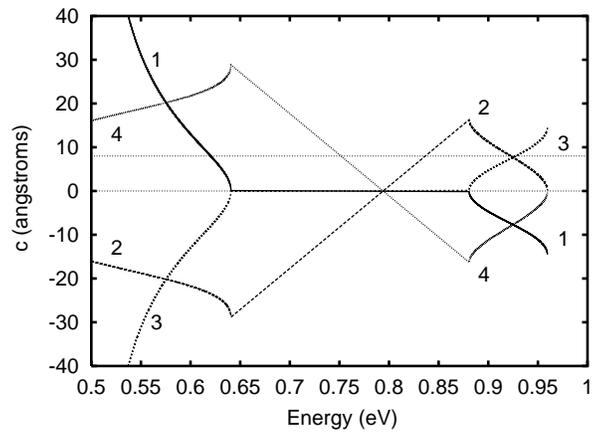}
%\fbox{\includegraphics[bb=0 0 8in 6in,scale=0.5]{cbsfig3.eps}}
\end{center}
\caption{Same as Fig. 4 but for a central barrier. (The 
straight lines across the allowed zone connect related solutions.)}
 \label{fig7} 
\end{figure}

\begin{figure}[htb]
\begin{center}
\leavevmode
\epsfxsize=8cm
\begin{tabular}{|cc|} \hline
 \epsffile{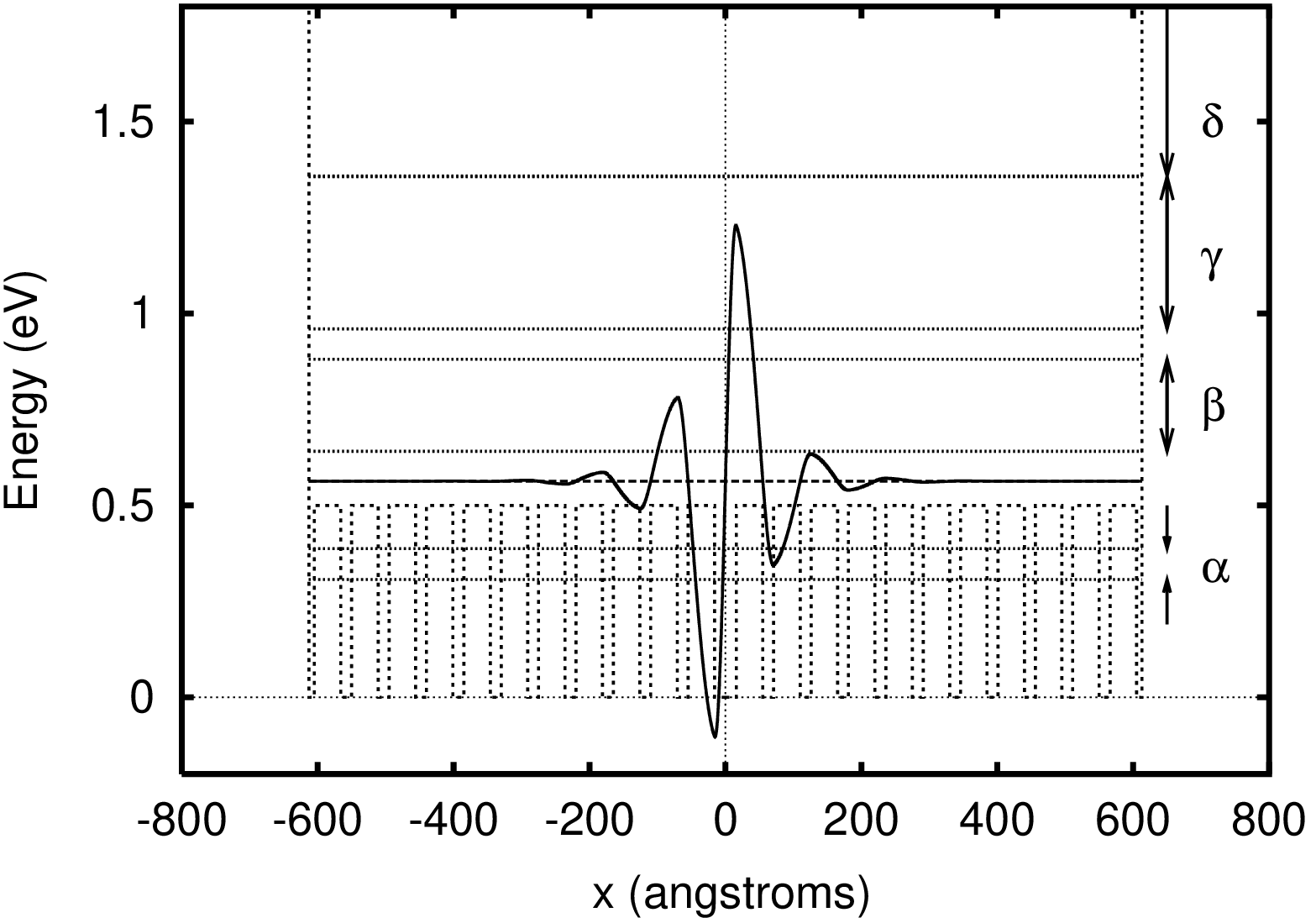} &
\epsfxsize=8cm \epsffile{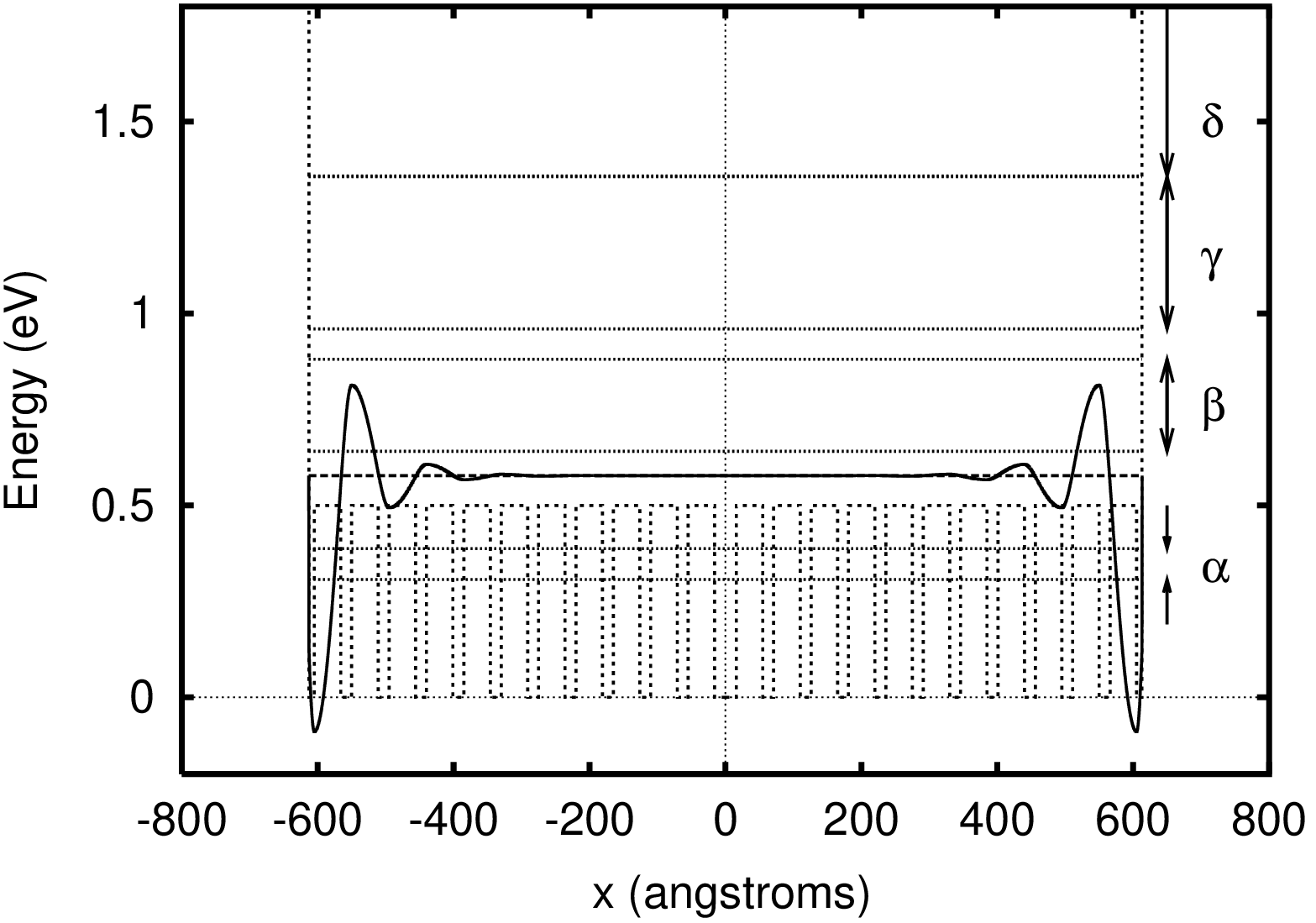} \\ \hline
a) & b) \\ \hline
\epsfxsize=8cm \epsffile{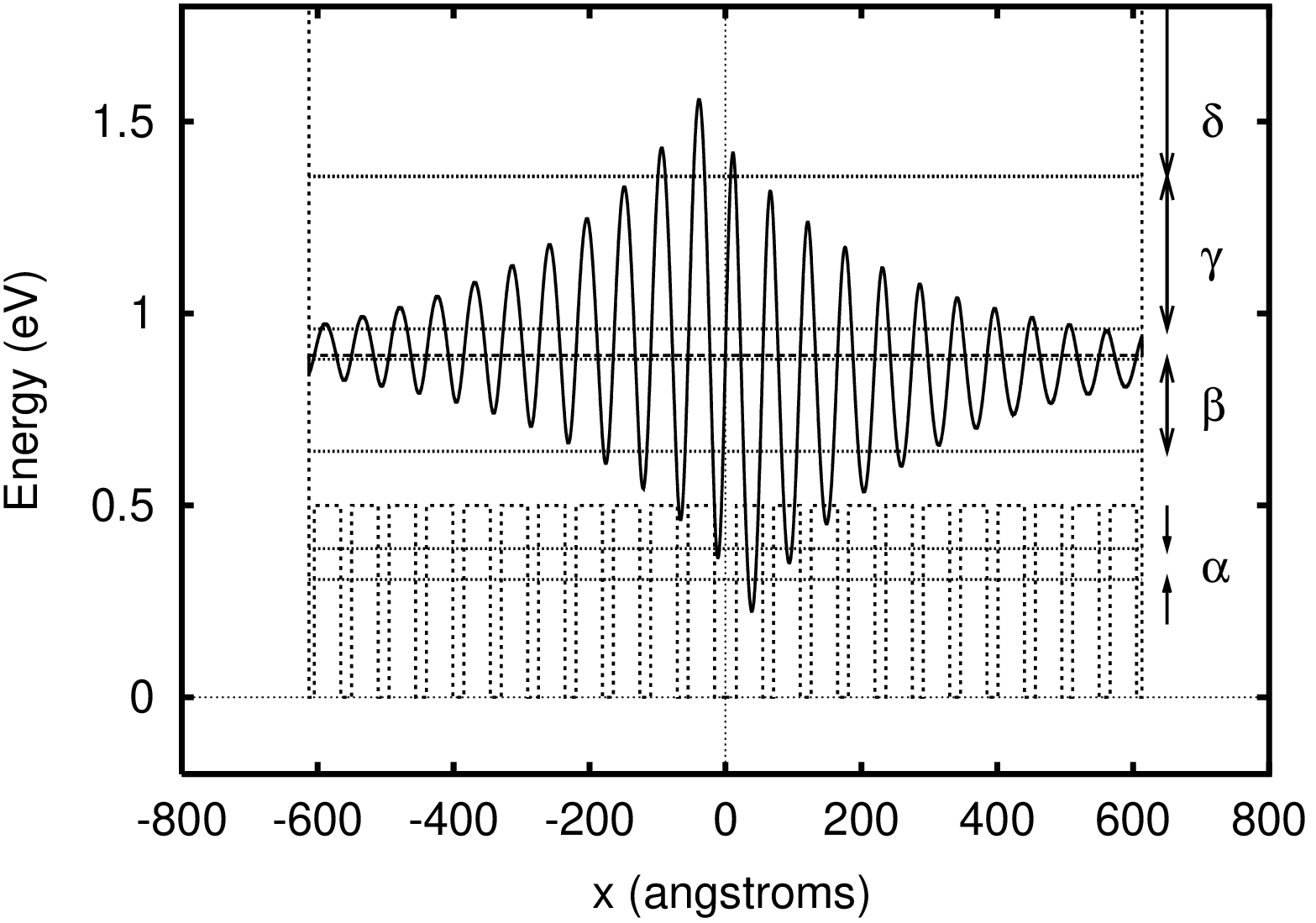} &
\epsfxsize=8cm \epsffile{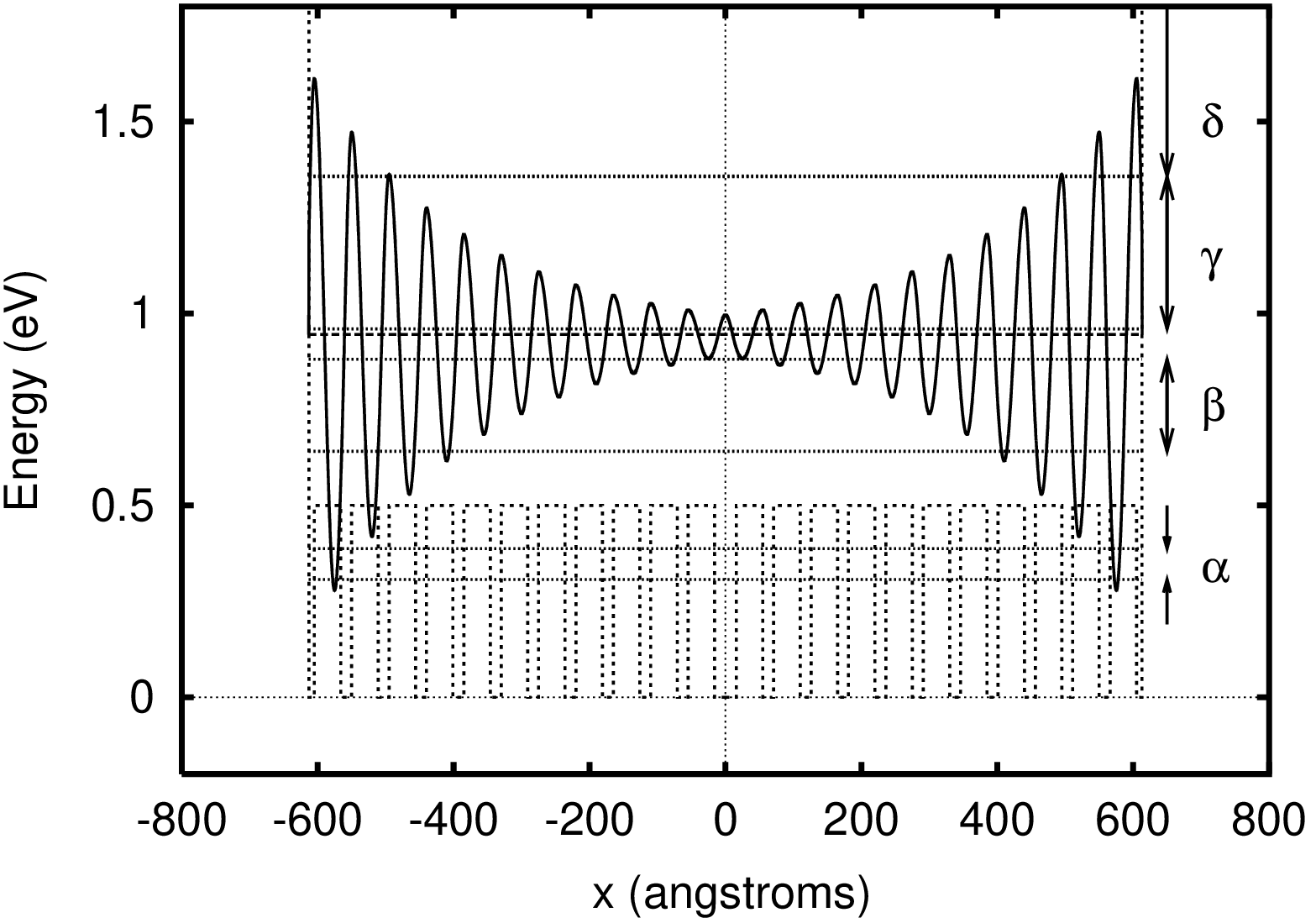} \\ \hline
c) & d) \\ \hline
\end{tabular}
\end{center}
\caption{States in a box, for c=8\AA, also showing a) the CBS at 563 
meV, b) the ABS at 577 meV, c) the CBS at 891 meV and d) the ABS at 946 meV. 
At right, $\alpha$ labels the first allowed band from 307$\to$387 meV, $\beta$
the second allowed band from 641$\to$881 meV, $\gamma$ the third allowed band 
from 960$\to$1357 meV, and $\delta$ the fourth allowed band from 1357 
meV. Dotted lines show the potential cells, in eV; wave functions 
are dimensionless and are drawn with base line at the energy 
eigenvalue.}
 \label{fig8} 
\end{figure}

\begin{figure}[htb]
\begin{center}
\leavevmode
\epsfxsize=8cm
\begin{tabular}{|cc|} \hline
a) & \epsffile{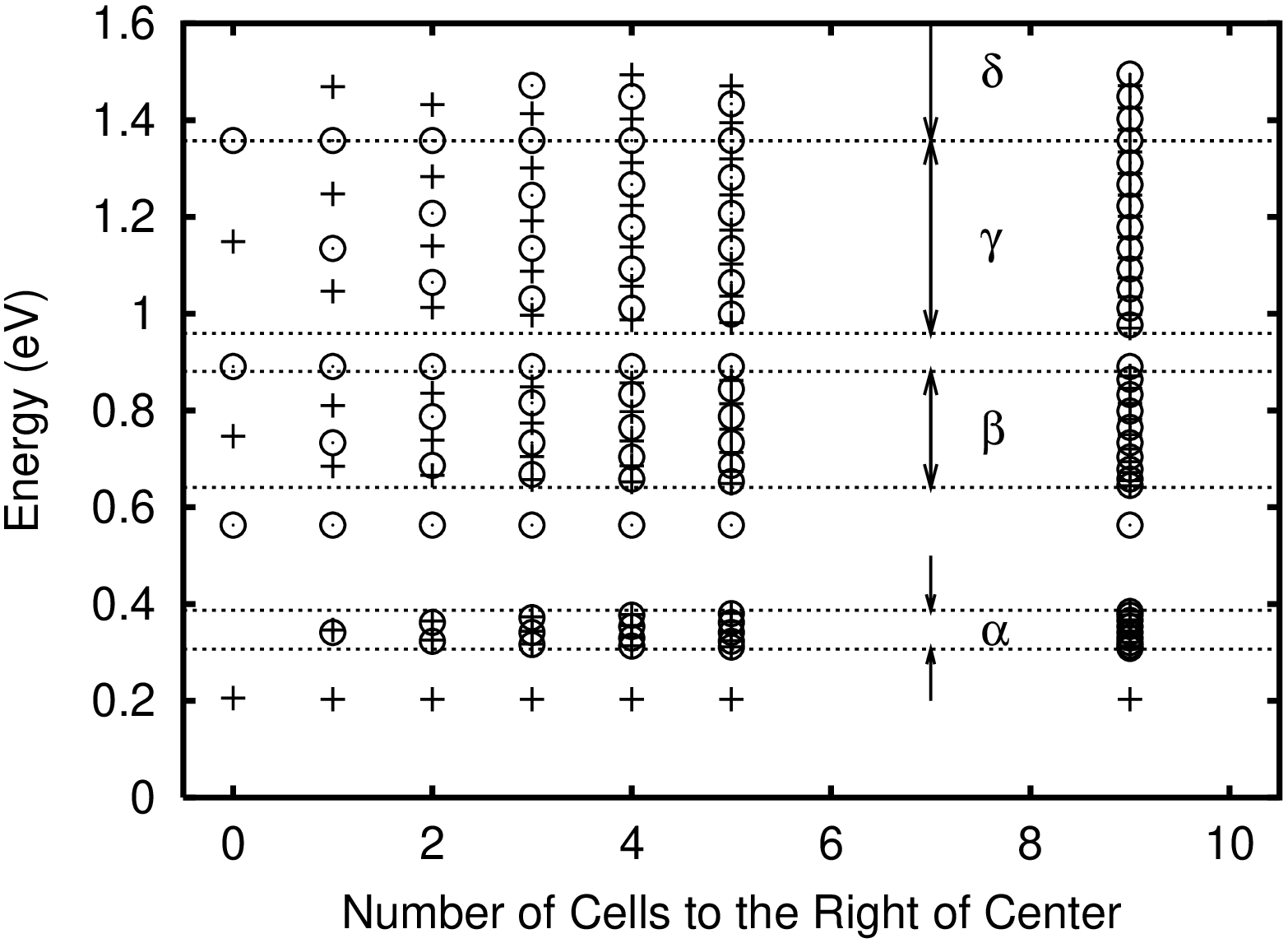} \\ \hline
b) & \epsfxsize=8cm\epsffile{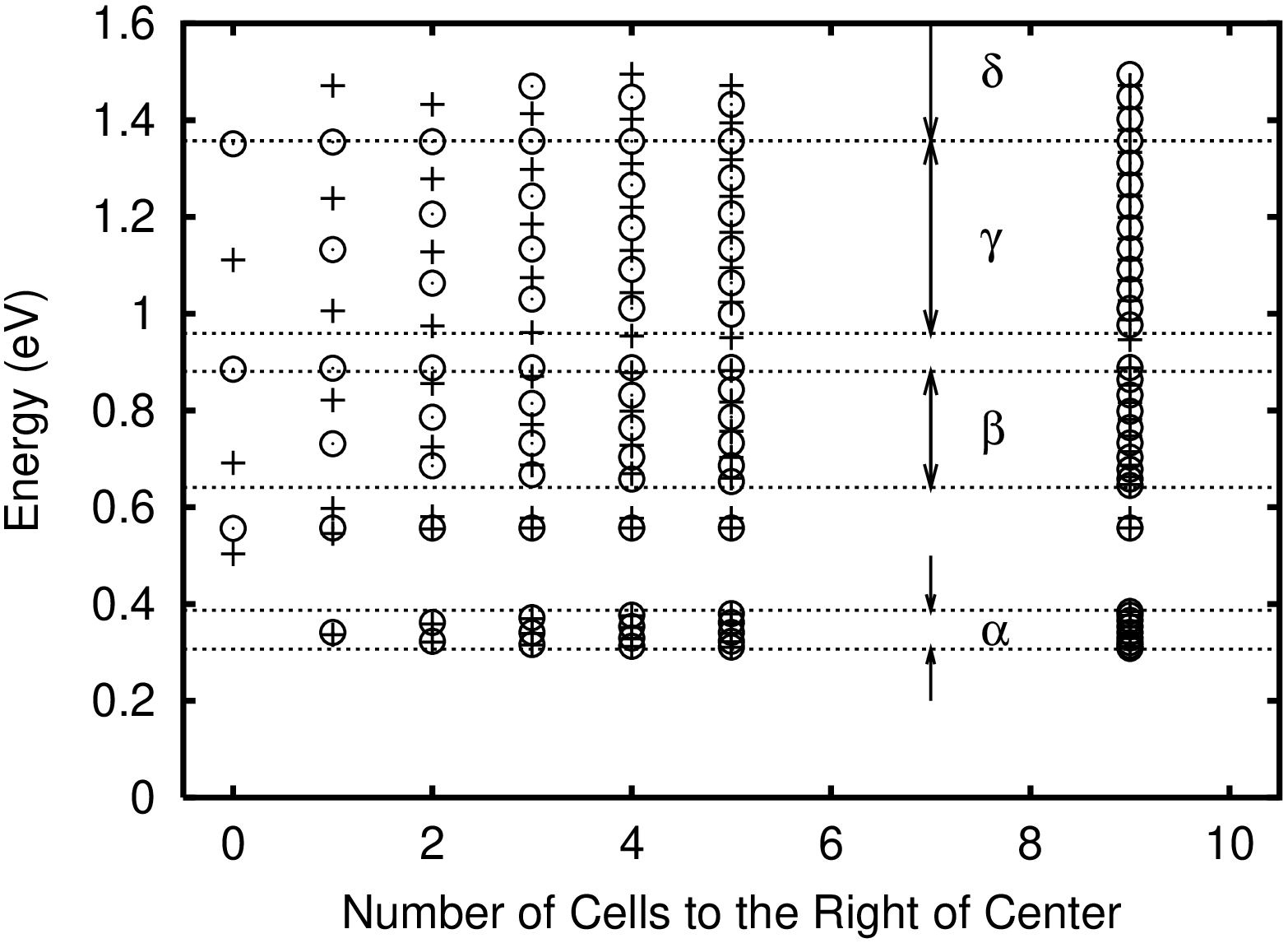} \\ \hline
%a) & \includegraphics[bb= 0 0 8in 6in,scale=0.5]{energyeo.eps} \\ \hline
%b) & \includegraphics[bb=0 0 8in 6in,scale=0.5]{benergyeo.eps} \\ \hline
\end{tabular}
\end{center}
\caption{Even (+) and odd (O) state energies (in a box) versus number 
$N$ of cells to the right of the central defect. Panel a) is for a 
central well, and b) a central barrier. Bands labelled as in Fig. 8. 
} 
 \label{fig9} 
\end{figure}

\begin{figure}[htb]
\begin{center}
\leavevmode
\epsfxsize=8cm
\epsffile{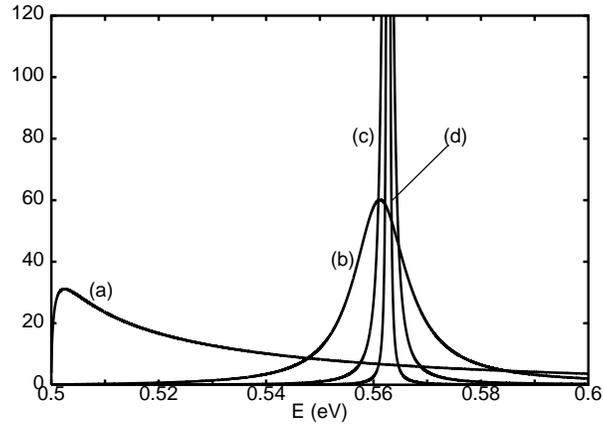}    %{cbsfig10.eps}    
%\fbox{\includegraphics[bb=0 0 8in 6in,scale=0.5]{intgraph.eps}}
\end{center}
\caption{The transition strength times density of states (units 
ev$^{-1} \AA^{-2}$) for a) a central well of width $32 \AA$, and a 
central well surrounded by b) one cell, c) two cells, and d) three 
cells on each side.} 
 \label{fig10}
\end{figure}

\begin{figure}[htb]
\begin{center}
\leavevmode
\epsfxsize=8cm
\begin{tabular}{|cc|} \hline
 \epsffile{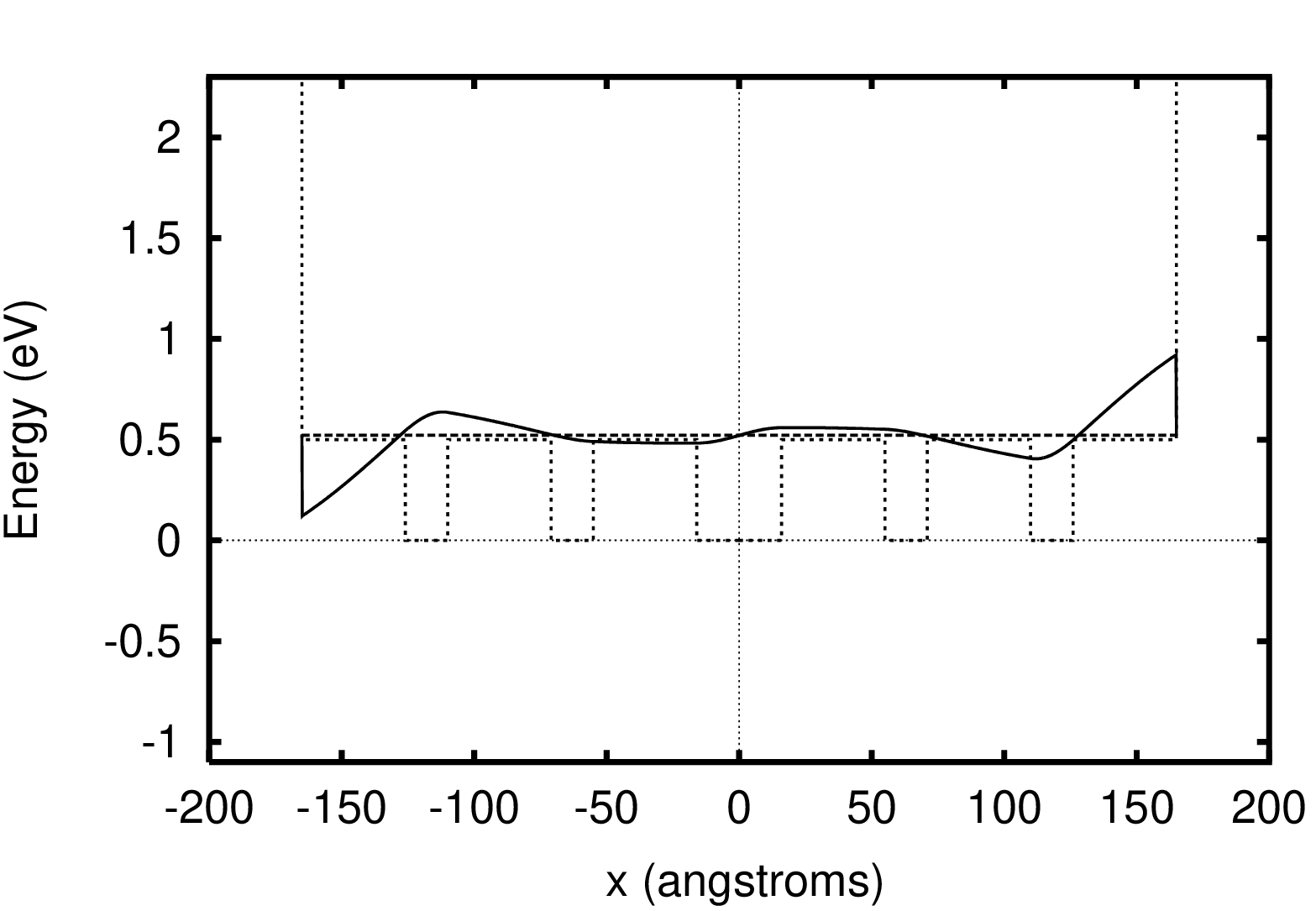} &
\epsfxsize=8cm \epsffile{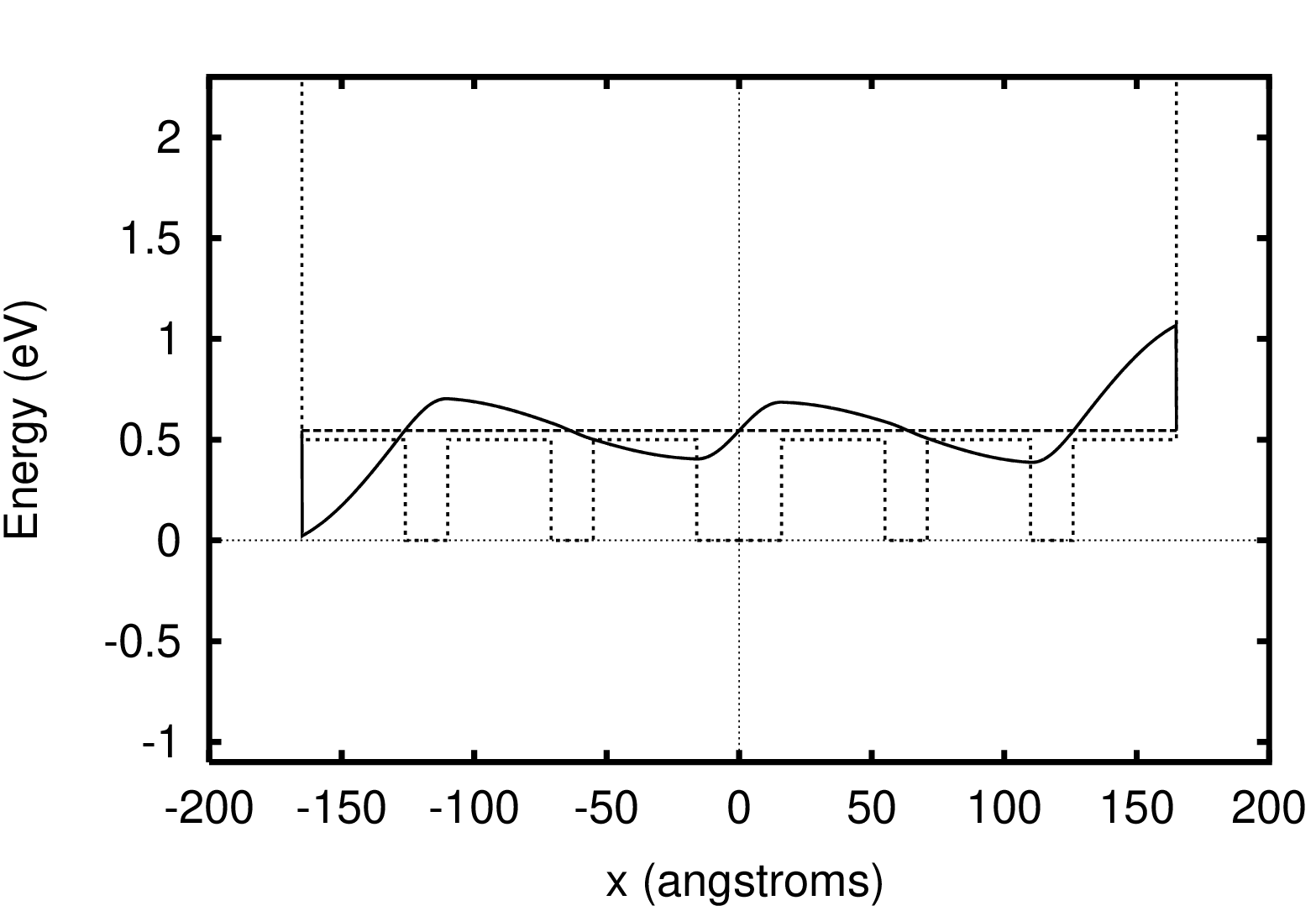} \\ \hline
a) & b) \\ \hline
\epsfxsize=8cm \epsffile{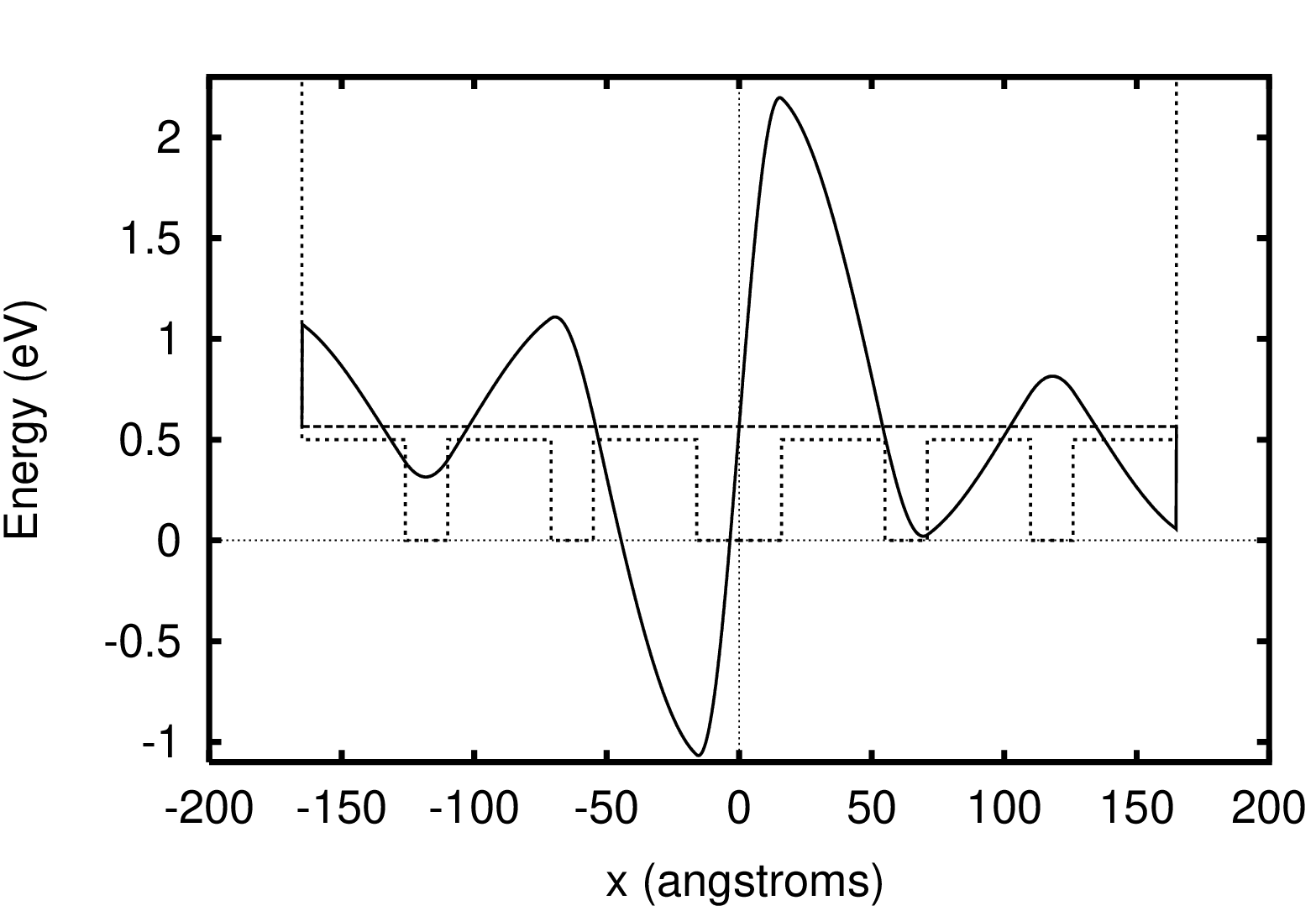} &
\epsfxsize=8cm \epsffile{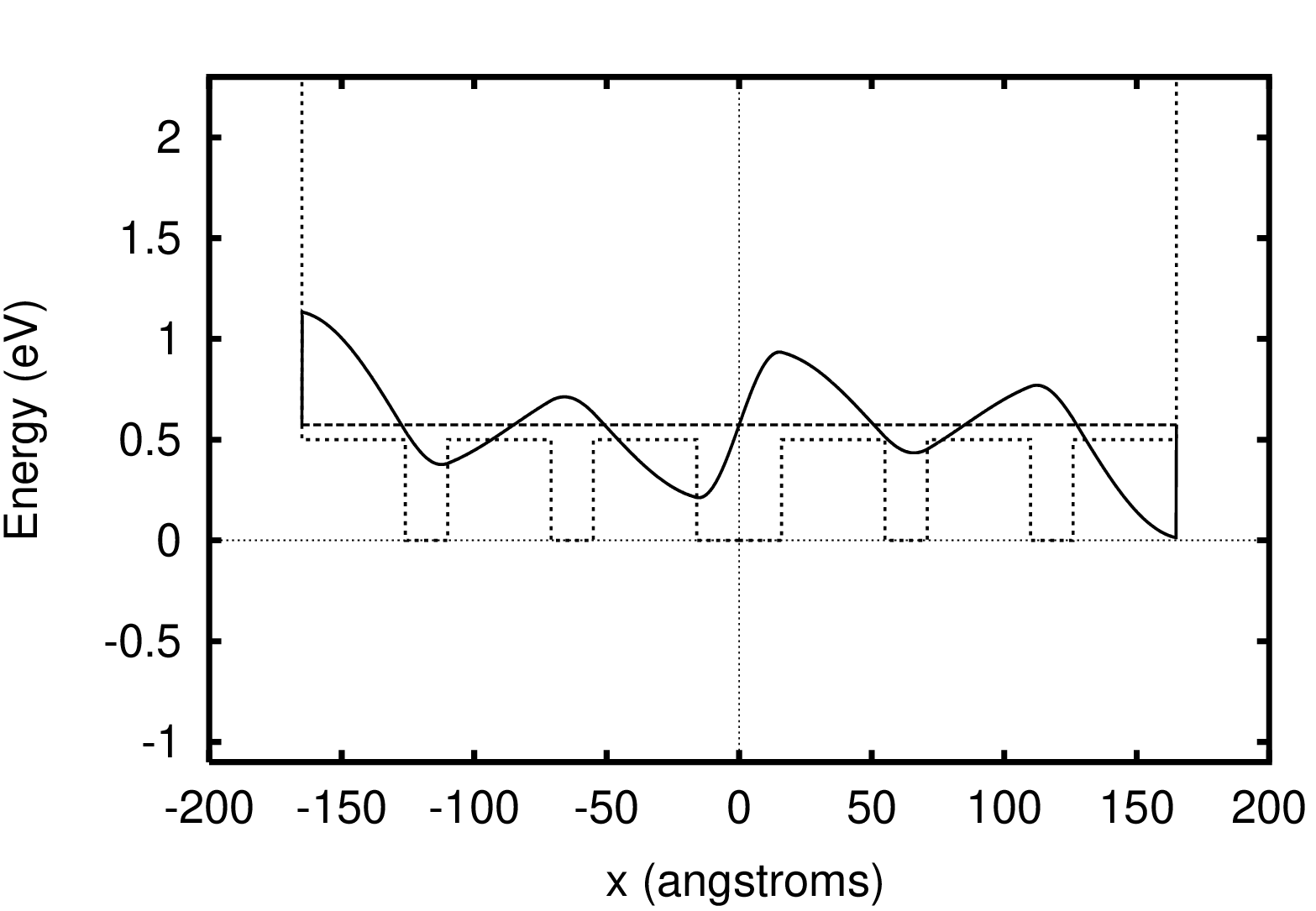} \\ \hline
c) & d) \\ \hline
\end{tabular}
\end{center}
\caption{Four representative continuum wave functions, for c=8\AA, 
at energies $E = 521$ meV (a); $545$ meV (b); $565$ meV (c); and 
$573$ meV (d). Dotted lines show the potential cells, in eV; wave functions 
are dimensionless and are drawn with base line at the corresponding 
energy.}
 \label{fig12} 
\end{figure}

\begin{figure}[htb]
\begin{center}
\leavevmode
\epsfxsize=8cm
\epsffile{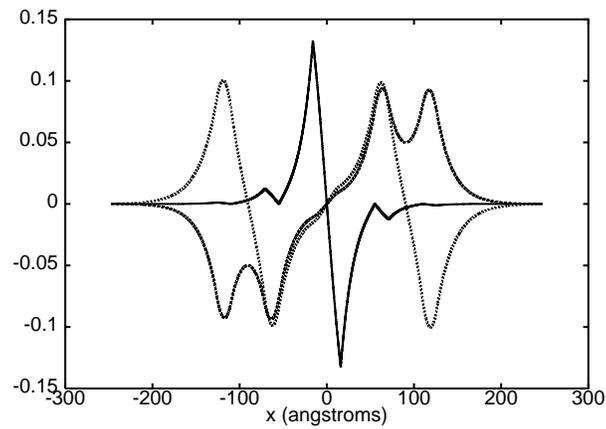}      %{cbsfig11.eps}      
%\fbox{\includegraphics[bb=0 0 8in 6in,scale=0.5]{3waves.eps}}
\end{center}
\caption{The derivative of the ground state wave function (over 
$m^*$) a) and the first and second odd-parity excited wave functions, 
b) and c), for a 5 well potential (two identical cells on each side of 
a central well), enclosed in a box, illustrating the similar overlap 
near the origin.} 
 \label{fig11}
\end{figure}

\begin{figure}[htb]
\begin{center}
\leavevmode
\epsfxsize=8cm
\begin{tabular}{|cc|} \hline
$N=1$ & \epsffile{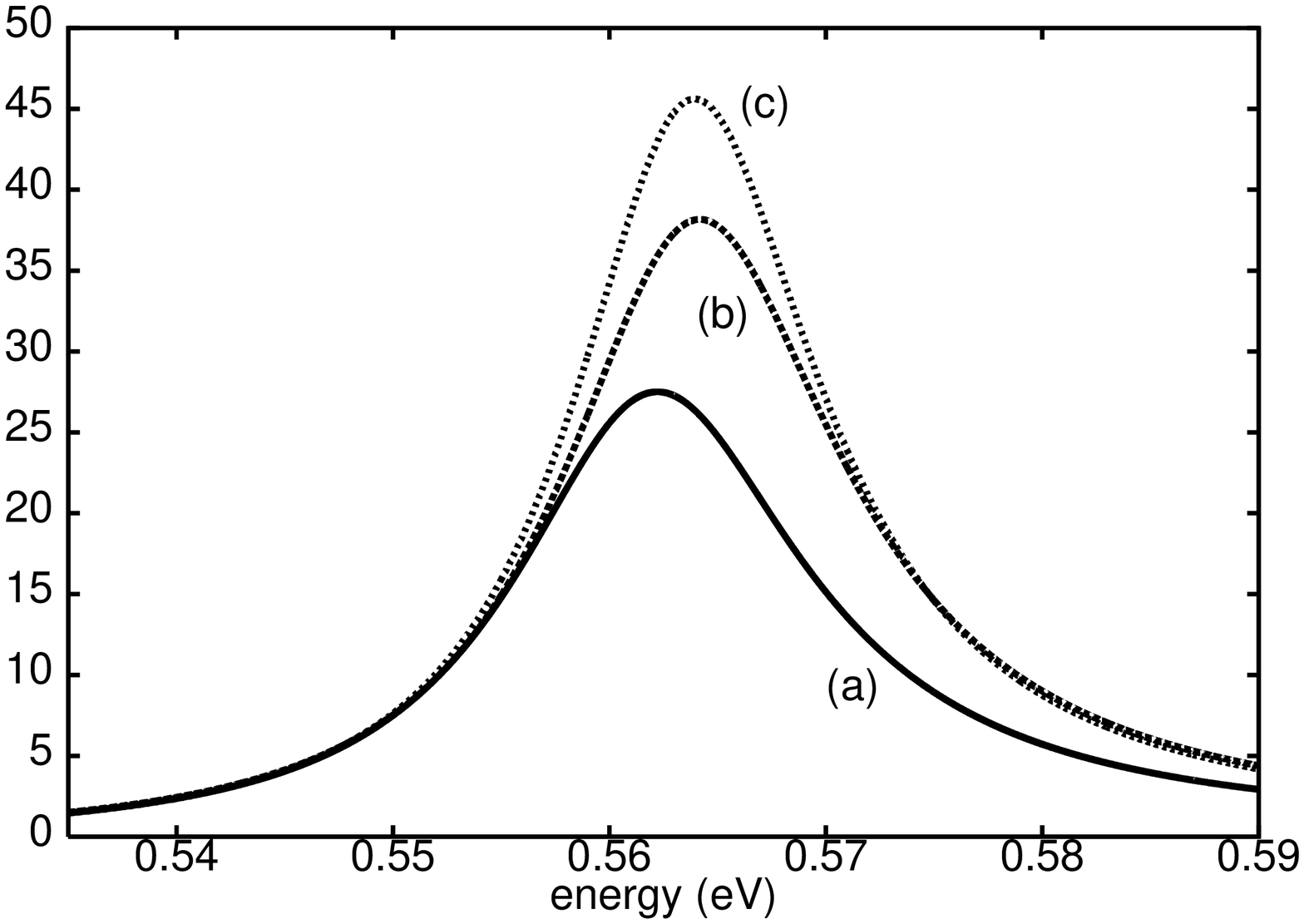} \\ \hline
$N=2$ & \epsfxsize=8cm \epsffile{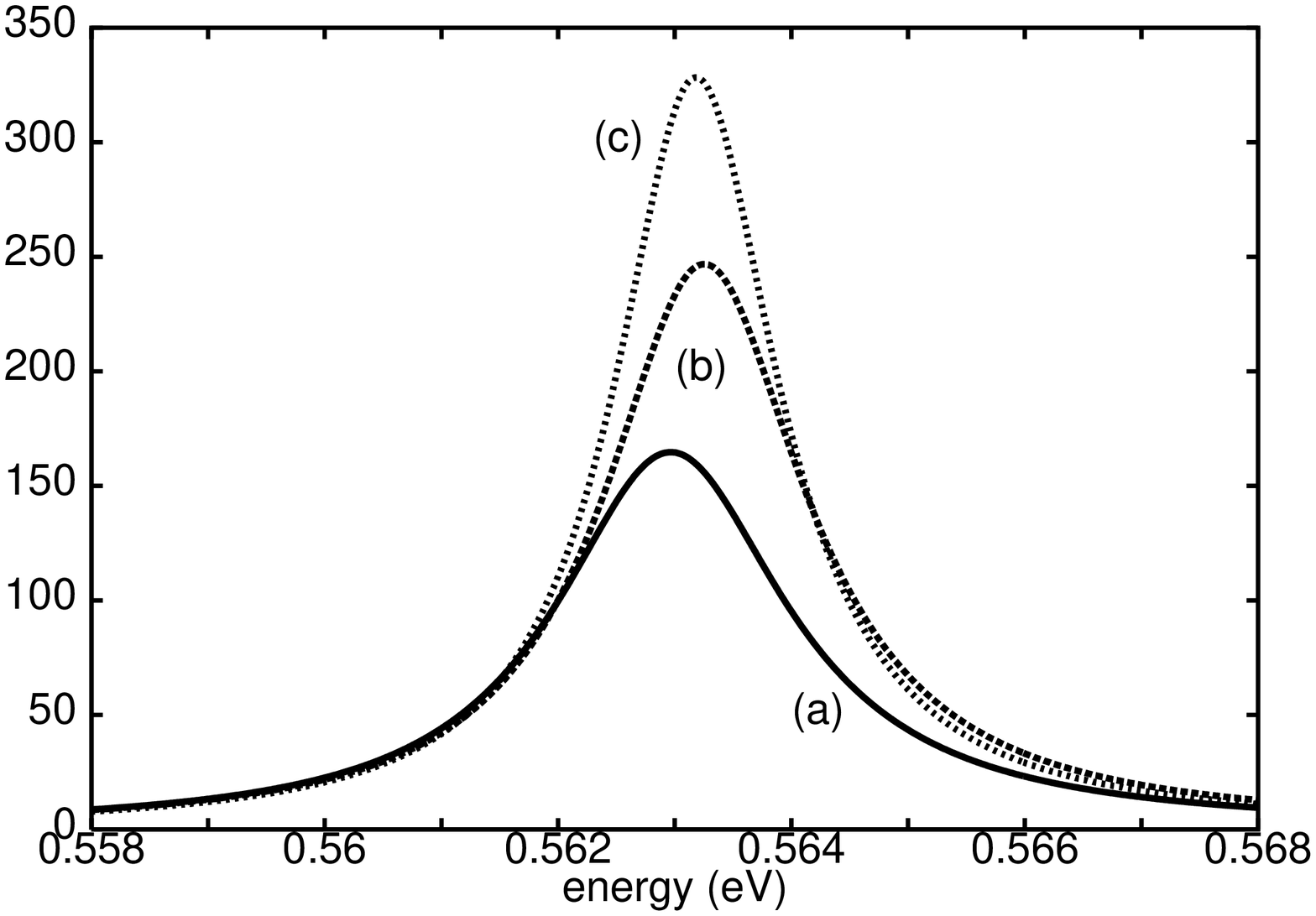} \\ \hline
%$N=1$ & \includegraphics[bb= 0 0 8in 6in,scale=0.5]{cbsfg13a.ps} \\ \hline
%$N=2$ & \includegraphics[bb=0 0 8in 6in,scale=0.5]{cbsfg13b.ps} \\ \hline
\end{tabular}
\end{center}
\caption{Transition strength to continuum (units ev$^{-1} \AA^{-2}$) 
for (a): case $(Q,Q)$, (b): $(0,H)_1$, and (c): $(0,H)_2$, showing 
the evolution of the CBS peak. Note change of scale between the upper 
panel for $N = 1$ reflector and the lower panel for two. } 
 \label{fig13} 
\end{figure} 

\end{onecolumn}

%\vfill\noindent file: c:/dd/n01/cbs01/cbsf.tex
\end{document}